\documentclass[a4paper,12pt]{elsarticle}

\usepackage[top=0.5in,bottom=0.75in,left=0.6in,right=0.6in]{geometry}
\usepackage{graphicx}
\usepackage[colorlinks=true,linkcolor=blue,citecolor=blue,urlcolor=blue]{hyperref}
\usepackage[figurename=Fig.]{caption}
\usepackage{color}
\usepackage{float}
\usepackage{xcolor}
\usepackage{amssymb}
\usepackage{amsmath}
\usepackage{mathrsfs}
\usepackage{upgreek}
\usepackage{makecell}
\usepackage{cases}
\usepackage{pdflscape}
\usepackage{subfigure}
\usepackage{pdfpages}
\usepackage{lineno}
\usepackage{booktabs}
\usepackage{makeidx}
\usepackage{bm}
\usepackage{xstring}
\usepackage{epstopdf}
\usepackage{lipsum}
\usepackage{slashed}
\usepackage{multicol}
\usepackage{setspace}

\usepackage[separate-uncertainty]{siunitx}

\biboptions{numbers,sort&compress}

\newcommand{\doiref}[2]{\href{http://dx.doi.org/#1}{#2}}
\newcommand{\dsl}[2]{{\slashed #1}\!_{#2}}
\newcommand{\To}[1][-0.25]{\hspace{#1em}\to\hspace{#1em}}
\newcommand{\e}[1]{\times 10^{#1}}
\newcommand{\er}[3]{^{-#1}_{+#2}\times 10^{#3}}
\definecolor{myblue}{RGB}{65,111,166}
\definecolor{myred}{RGB}{168,66,63}
\definecolor{mygreen}{RGB}{134,164,74}
\definecolor{mypurple}{RGB}{110,84,141}
\definecolor{myindigo}{RGB}{61,150,174}
\definecolor{myorange}{RGB}{218,129,55}
\definecolor{mylightblue}{RGB}{142,165,203}

\makeatletter
\def\ps@pprintTitle{%
  \let\@oddhead\@empty
  \let\@evenhead\@empty
  \let\@oddfoot\@empty
  \let\@evenfoot\@oddfoot
}
\makeatother



\begin{document}

%\vspace{5cm}

\begin{frontmatter}

\title{\ \\$D$-Wave Charmonia $\eta_{c2}(1^1\!D_2)$, $\psi_2(1^3\!D_2)$, and $\psi_3(1^3\!D_3)$ in $B_c$ Decays}

\author[]{Qiang Li}\ead{lrhit@protonmail.com}
\author[]{Tianhong Wang\corref{corauthor}}
\ead{thwang@hit.edu.cn}
\cortext[corauthor]{Corresponding author}
\author[]{Yue Jiang}\ead{jiangure@hit.edu.cn}
\author[]{Han Yuan}\ead{hanyuan@hit.edu.cn}
\author[]{Guo-Li Wang}\ead{gl\_wang@hit.edu.cn}

\address{Department of Physics, Harbin Institute of Technology, Harbin, 150001, P. R. China}

\begin{abstract}
We study the semi-leptonic and non-leptonic decays of $B_c$ meson to $D$-wave charmonia, namely, $\eta_{c2}(1^1\!D_2)$, $\psi_2(1^3\!D_2)$, and $\psi_3(1^3\!D_3)$. In our calculations, the instantaneous Bethe-Salpeter method is applied to achieve the hadronic matrix elements. This method includes relativistic corrections which are important especially for the higher orbital excited states. For the semi-leptonic decay channels with electron as the final lepton, we get the  branching ratios $\mathcal{B}[B_c \To \eta_{c2}e\bar{\nu}_e] = 5.9\er{0.8}{1.0}{-4}$, $\mathcal{B}[B_c \To \psi_2e\bar{\nu}_e]=1.5\er{0.2}{0.3}{-4}$, and $\mathcal{B}[B_c \hspace{-0.3em} \to \hspace{-0.3em}\psi_3e\bar{\nu}_e]=3.5\er{0.6}{0.8}{-4}$. The transition form factors, forward-backward asymmetries, and lepton spectra in these processes are also presented. For the non-leptonic decay channels, those with $\rho$ as the lighter meson have the largest branching ratios, $\mathcal{B}[B_c \To \eta_{c2}\rho] = 8.1\er{1.0}{1.0}{-4}$, $\mathcal{B}[B_c \To \psi_2\rho]=9.6\er{1.0}{1.0}{-5}$, and $\mathcal{B}[B_c \To \psi_3\rho]=4.1\er{0.7}{0.8}{-4}$.
\end{abstract}

\end{frontmatter}

%\linenumbers

\section{Introduction}

In 2013, the Belle Collaboration reported the evidence of a new resonance $X(3823)$ in the $B$ decay channel $B^\pm\To X(\hspace{0.3em}\To\chi_{c1}\gamma)K^\pm$ with a statistical significance of 3.8$\sigma$~\cite{Belle-2013}. And very recently, the BESIII collaboration verified its existence with a statistical significance of $6.2\sigma$~\cite{BESIII-2015}. Both groups got the similar mass and the ratio of partial decay width for this particle. On one hand, this state has a mass of $3821.7 \pm 1.3(\rm{stat}) \pm 0.7(\rm{syst})$ MeV$/c^2$, which is very near the mass value of the $1^3\!D_2$ charmonium predicted by potential models~\cite{GI, PRD67-2003}; on the other hand, the electromagnetic decay channels $\chi_{c1}\gamma$ and $\chi_{c2}\gamma$ are observed while the later one is suppressed, which means the $1{^1\!D_2}$ and $1{^3\!D_3}$ charmonia cases are excluded.

To confirm the above experimental results and compare with other theoretical predictions, studying the properties of $D$-wave charmonia in a different approach is deserved. In this work we study the $\psi_2({1^3\!D_2})$ and its two partners $\eta_{c2}(1^1\!D_2)$ and $\psi_3(1^3\!D_3)$ in the weak decays of $B_c$ meson which has attracted lots of attention since its discovery by the CDF Collaboration at Fermilab~\cite{CDF-1998}. Unlike the charmonia and bottomonia which are hidden-flavor bound states, the $B_c$ meson, which consists of a bottom quark and a charm quark, is open-flavor. Besides that, it's the ground state, which means it cannot decay through strong or electromagnetic interaction. So the $B_c$ meson provides an ideal platform to study the weak interaction.

The semi-leptonic and non-leptonic transitions of the $B_c$ meson into charmonium states are important processes. Experimentally, only those with $J/\psi$ or $\psi(2S)$ as the final charmonium have been detected~\cite{PDG-2014}. As the LHC accumulates more and more data, the weak decay processes of $B_c$ meson to charmonia with other quantum numbers will have more possibilities to be detected. That is to say, this is an alternative way to study the charmonia, especially those have not yet been discovered, such as $\eta_{c2}(1^1\!D_2)$ and $\psi_3(1^3\!D_3)$. Theoretically, the semi-leptonic and non-leptonic transitions of the $B_c$ meson into $S$-wave charmonium states are studied widely by several phenomenological models, such as the relativistic constituent quark model~\cite{cch1,PRD71-2005,PRD73-2006,jpvary,korner,ebert1}, the non-relativistic constituent quark model~\cite{PRD74-2006}, the technique of hard and soft factorization~\cite{NPP-2002} and QCD factorization~\cite{dds2}, QCD sum rules~\cite{SR}, Light-cone sum rules~\cite{huang}, the perturbative QCD approach~\cite{dds1,pQCD,zhou-2014,zhou-2015}, and NRQCD~\cite{qiao1,qiao2}. There are also some theoretical models to study the processes of $B_c$ decay to a $P$-wave charmonium~\cite{PRD65-2001,PRD71-2005,lu1,lu2,PRD82-2010,NPP39-2012}, while we lack the information of $B_c$ decay to a $D$-wave charmonium.

Here we will use the Bethe-Salpeter (BS) method to investigate the exclusive semi-leptonic and non-leptonic decays of the $B_c$ meson to the $D$-wave charmonium. This method has been used to study processes with $P$-wave charmonium~\cite{PRD65-2001, NPP39-2012}. As is known to all, the BS equation~\cite{PR84-1951} is a relativistic two-body bound state equation. To solve BS equation of $D$-wave mesons and get corresponding wave function and mass spectra, we use the instantaneous approximation, that is, we solve the Salpeter equations~\cite{PR87-1952} which has been widely used for bound states decay problems~\cite{PLB633-2006,PLB674-2009,JHEP03-2013,3D23-wave}.

This paper is organized as follows. In \autoref{Sec-2} we present the general formalism for semi-leptonic and
non-leptonic decay widths of $B_c$ into $D$-wave charmonia. In \autoref{Sec-3} we give the analytic expressions of the corresponding form factors given by the BS method. In \autoref{Sec-4}, the numerical results are achieved and we compare our results with others', also the theoretical uncertainties and lepton spectra are presented in this section. \autoref{Sec-5} is a short summary of this work. Some bulky analytical expressions are presented in the Appendix.

\section{Formalisms of Semileptonic and Nonleptonic Decays}\label{Sec-2}
In this Section we will derive the general formalism for the calculations of both semi-leptonic and non-leptonic decay widths of $B_c$ meson.
\subsection{The Semi-leptonic Decay}
The semi-leptonic decays of $B_c$ meson into $D$-wave charmonia are three-body decay processes. We consider the neutrinos as massless fermions. The differential form of the three-body decay width can be written as
\begin{equation} \label{3-body-width}
\mathrm{d} \Gamma=\frac{1}{(2\uppi)^3}\frac{1}{32M^3}\overline{|\mathcal{M}|^2} \mathrm{d}m_{12}^2 \mathrm{d}m_{23}^2,
\end{equation}
where $M$ is the mass of $B_c$; $m_{12}$ is the invariant mass of final $c\bar c$ meson and neutrino which is defined as $m_{12}^2=(P_F+p_\nu)^2$; $m_{23}$ is the invariant mass of final neutrino and charged lepton, which is defined as $m_{23}^2=(p_\nu+p_\ell)^2$. Here we have used $P_F$, $p_\nu$ and $p_\ell$ to denote the 4-momentum of final $c\bar c$ meson, neutrino, and charged lepton, respectively. $\mathcal{M}$ is the invariant amplitude of this process. In above equation we have summed over the polarizations of final states. %and averaged over the polarizations of initial meson.

\subsubsection{Form Factors}
\begin{figure}[htbp]
\centering
\includegraphics[width = 0.5\textwidth]{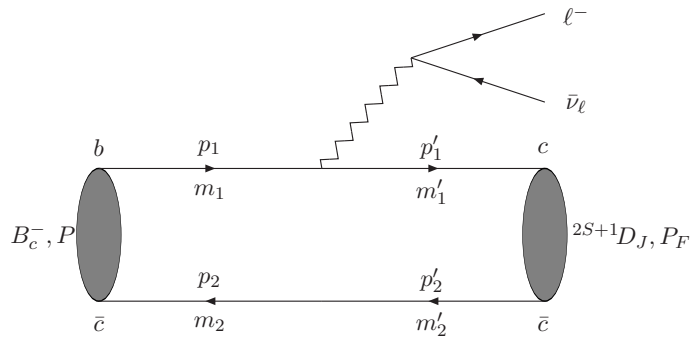}
\caption{Feynman diagram of the semi-leptonic decay of $B_c$ into $D$-wave charmonia. $P$ and $P_F$ are the  momenta of initial and final mesons, respectively. $S$, $D$, and $J$ are quantum numbers of spin, orbital angular momentum and total angular momentum for the final $c\bar{c}$ system, respectively.}\label{Pic-1}
\end{figure}

The Feynman diagram involved in the semi-leptonic decays of $B_c$ meson in the tree level is showed in \autoref{Pic-1}.
The invariant amplitude $\mathcal{M}$ can be written directly as
\begin{equation} \label{eq-amp1}
\mathcal{M}= \frac{G_F}{\sqrt{2}} V_{cb} \langle c\bar c |h^{\mu}|B_c \rangle \bar{u}_{\ell}({p_\ell}) \Gamma_{\mu} v_{\nu}({p_\nu}) ,
\end{equation}
where $G_{\!F}$ is the Fermi constant; $V_{cb}$ is the CKM matrix element for $b\To c$ transition; $\langle c\bar c |h^{\mu}|B_c \rangle$ is the hadronic matrix element; $h^{\mu}=\bar{c} \Gamma^{\mu}{b}$ is the weak charged current and $\Gamma^\mu=\gamma^\mu(1-\gamma^5)$.
The general form of the hadronic matrix element $\langle c \bar c |h^{\mu}|B_c \rangle$ depends on the total angular momentum $J$ of the final meson. For $\eta_{c2}$, $J=2$, the transition matrix can be written as
\begin{equation} \label{form-1}
\langle c\bar c |h^{\mu}|B_c \rangle=e_{\alpha\beta}P^{\alpha}(s_1P^{\beta}P^{\mu}+s_2P^{\beta}P_F^{\mu}+s_3g^{\beta\mu}+\mathrm{i}s_4\epsilon^{\mu\beta PP_F}),
\end{equation}
where $g^{\beta\mu}$ is the Minkowski metric tensor.
We have used the definition $\epsilon_{\mu \nu P P_F}\equiv\epsilon_{\mu \nu \alpha \beta }P^{\alpha} P^{\beta}_F$; $\epsilon_{\mu \nu \alpha \beta}$ is the totally antisymmetric tensor; $e_{\alpha\beta}$ is the polarization tensor of the charmonium with $J=2$; $s_1, s_2, s_3$ and $s_4$ are the form factors for $^1\!D_2$ state; for ${^3}\!{D_2}$ state the relation between $\langle c\bar c |h^{\mu}|B_c \rangle$ and form factors $t_i~(i=1,2,3,4)$ has the same form with $^1\!D_2$ just $s_i$ replaced with $t_i$.
For the $J=3$ meson, the hadronic matrix element can be described by form factors $h_i$ as below
\begin{equation}\label{form-3}
\langle c\bar c |h^{\mu}|B_c \rangle=e_{\alpha\beta\gamma}P^{\alpha}P^{\beta} (h_1P^{\gamma}P^{\mu}+h_2P^{\gamma}P_F^{\mu}+h_3
g^{\gamma\mu}+\mathrm{i}h_4\epsilon^{\mu \gamma PP_F}),
\end{equation}
where $e_{\alpha\beta\gamma}$ is the polarization tensor for the meson with $J=3$. The expressions of these form factors are given in the next section.

The squared transition matrix element with the summed polarizations of final states (see Eq.~(\ref{3-body-width})) has the form
\begin{equation} \label{eq-Squaramp}
\overline{|\mathcal{M}|^2} =\frac{G_F^2}{2} |V_{cb}|^2 L^{\mu \nu} H_{\mu\nu}.
\end{equation}
In the above equation $L^{\mu \nu}$ is the leptonic tensor
\begin{equation} \label{eq-LT}
\begin{aligned}
L^{\mu \nu}=& \sum_{s_\ell,s_\nu}[\bar{u}_{\ell}({p_\ell}) \Gamma^{\mu} v_{\nu}({p_\nu})][\bar{u}_{\ell}({p_\ell}) \Gamma^{\nu} v_{\nu}({p_\nu})]^\dag\\
           =& 8(p_\ell^{\mu}p_\nu^{\nu}+p_\nu^{\mu}p_\ell^{\nu}-p_\ell \! \cdot \! p_\nu g^{\mu \nu}-\mathrm{i}\epsilon^{\mu \nu p_\ell p_\nu}),
\end{aligned}
\end{equation}
and $H_{\mu \nu}$ is the hadronic tensor which can be written as
\begin{gather} \label{eq-HT}
H_{\mu \nu}= N_{1}P_{\mu}P_{\nu}+N_2(P_{\mu}{P_F}_\nu+P_{\nu}{P_F}_\mu)+N_4P_{F\mu}P_{F\nu}+
N_5g_{\mu \nu}+\mathrm{i} N_6 \epsilon_{\mu \nu P P_F},
\end{gather}
where $N_i~(i=1,2,4,5,6)$ is described by form factors $s_j$, $t_j$ or $h_j~(j=1,2,3,4)$~(see \ref{Ni}).
By using Eq.~(\ref{eq-LT}) and Eq.~(\ref{eq-HT}), we can write $L^{\mu\nu}H_{\mu\nu}$ as follow
\begin{equation}\label{LH}
\begin{aligned}
L^{\mu\nu}H_{\mu\nu}&=8 {N_1} \left(2 {P \!\cdot \! p_{\ell}} {P \!\cdot \! p_{\nu}  - M^2 {p_{\nu} \!\cdot \! p_{\ell}}}\right)
+16 {N_2} ({P \!\cdot \! p_{\ell} } {P_{F} \!\cdot \! p_{\nu} }+{P_{F} \!\cdot \! p_{\ell} } {P \!\cdot \! p_{\nu} }-{p_{\nu} \!\cdot \! p_{\ell}} {P\!\cdot \! P_F})\\
&+8 {N_4} \left(2 {P_{F} \!\cdot \! p_{\ell} } {P_{F} \!\cdot \! p_{\nu} } -M_F^2p_{\nu} \!\cdot \! p_{\ell}  \right)
-16 {N_5} {p_{\nu} \!\cdot \! p_{\ell}}
+16 {N_6} ({P_{F} \!\cdot \! p_{\ell} } {P \!\cdot \! p_{\nu} }-{P \!\cdot \! p_{\ell} } {P_{F} \!\cdot \! p_{\nu} }),
\end{aligned}
\end{equation}
where $M_F$ stands for the mass of final charmonium meson.
%\vspace{0.5em}
\subsubsection{Angular distribution and lepton spectra}

The angular distribution of semi-leptonic decays of $B_c$ to $D$-wave charmonia can be described as
\begin{equation}
\frac{\text{d} \Gamma}{ \text d\cos\theta} = \int \frac{1}{(2\uppi)^3} \frac{|\bm{p}^*_\ell||\bm{p}^*_F|}{16M^3} \overline{|\mathcal{M}|^2} \text{d} m^2_{23},
\end{equation}
where $\bm{p}^*_\ell$ and $\bm{p}^*_F$ are respectively the 3-momenta of the charged lepton and the final charmonium in the rest frame of lepton-neutrino system, which have the form $|\bm{p}^*_{\ell}|=\lambda^{\frac{1}{2}}(m^2_{23}, M_{\ell}^2, M^2_{\nu})/(2m_{23})$ and
$|\bm{p}^*_{F}|=\lambda^{\frac{1}{2}}(m^2_{23}, M^2, M^2_F)/(2m_{23})$. Here we used the K${\rm \ddot a}$llen function $\lambda(a,b,c)=(a^2+b^2+c^2-2ab-2bc-2ac)$. $M_{\ell}$ and $M_{\nu}$ are the masses of the charged lepton and neutrino, respectively. $\theta$ is angle between  $\bm{p}^*_\ell$ and $\bm{p}^*_F$. The forward-backward asymmetry $A_{FB}$ is another quantity we are interested, which is defined as
\begin{equation}
A_{FB} = \frac{\Gamma_{\cos\theta>0}-\Gamma_{\cos\theta<0}}{\Gamma_{\cos\theta>0}+\Gamma_{\cos\theta<0}}.
\end{equation}
One can check that $A_{FB}$ has the same value for the decays of $B_c^{+}$ and $B_c^-$ mesons. Its numerical results are given in \autoref{Sec-4}.
The momentum spectrum of charged lepton in the semi-leptonic decays is also an important quantity both experimentally and theoretically, which has the form
\begin{equation}
\frac{\text{d} \Gamma}{\text{d} |\bm{p}_\ell|} =\int \frac{1}{(2\uppi)^3} \frac{|\bm{p}_\ell|} {16M^2 E_\ell} \overline{|\mathcal{M}|^2} \text{d} m^2_{23},
\end{equation}
where $E_\ell$ is the energy of the charged lepton in the $B_c$ rest frame.

\subsection{Non-leptonic decay formalism}
\begin{figure}[htbp]
\centering
\includegraphics[width = 0.5\textwidth]{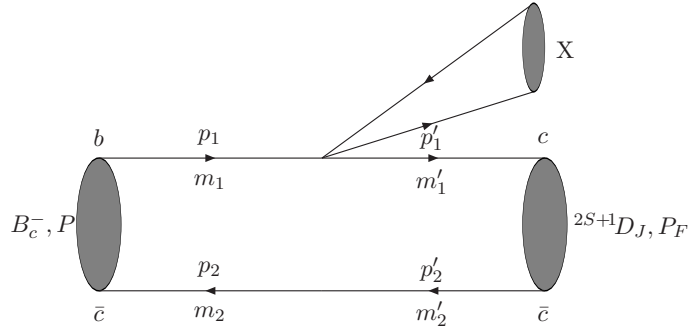}
\caption{The Feynman diagram of the non-leptonic decay of $B_c$ meson to $D$-wave charmonia. $X$ denotes a light meson.}\label{Pic-2}
\end{figure}

In this subsection, we will deal with the non-leptonic decays in the framework of factorization approximation~\cite{NPB133-1978, PL73-1978} .
The Feynman diagram of the non-leptonic decay of $B_c$ meson is showed in \autoref{Pic-2}. In this work we only calculate the processes when $X$ is $\pi$, $\rho$, $K$, or $K^*$.

The effective Hamiltonian for this process is~\cite{RMP68-1996}
\begin{equation}\label{H-eff}
H_{\text{eff}}=\frac{G_F}{\sqrt{2}} V_{cb}[c_1(\mu) O_1+c_2(\mu) O_2]+h.c.,
\end{equation}
where $c_1(\mu)$ and $c_2(\mu)$ are the scale-dependent Wilson coefficients. $O_i$s are the relevant four-quark local operators, which have the following forms
\begin{gather}
O_1=[V_{ud}(\bar d_{\alpha}u_{\alpha})_{V-A}+V_{us}(\bar s_{\alpha}u_{\alpha})_{V-A}](\bar c_{\beta} b_{\beta})_{V-A},\\
O_2=[V_{ud}(\bar d_{\alpha}u_{\beta})_{V-A} +V_{us}(\bar s_{\alpha}u_{\beta})_{V-A}] (\bar c_{\beta} b_{\alpha})_{V-A},
\end{gather}
where we have used the symbol $(\bar q_1 q_2)_{V-A}=\bar q_1 \gamma^{\mu}(1-\gamma^5) q_2$; here $\alpha$ and $\beta$ denote the color indices.

As a primary study, in this work the non-leptonic $B_c$ decays are calculated with the factorization approximation, which has been widely used in heavy mesons' weak decays~\cite{cch1,PRD73-2006,PRD74-2006,Faustov-2013}. In this approximation, the decay amplitude is factorized as the product of two parts, namely, the hadronic transition matrix element and an annihilation matrix element. The factorization assumption is expected to hold for process that involve a heavy meson and a light meson, provided the light meson is energetic~\cite{Dugan-1991}. Then we can write the non-leptonic decay amplitude as
\begin{equation} \label{nonleptonic amp}
\mathcal{M}\big [B_c\to(c\bar c)X \big ] \simeq \frac{G_F}{\sqrt{2}} V_{bc} V_{q_1 q_2} a_1(\mu) \langle c \bar c|h^{\mu}_{bc}|B_c \rangle \langle X|J_{\mu}|0\rangle.
\end{equation}
In above equation we have used the definitions $J_{\mu}=(\bar q_1 q_2)_{V-A}$; $a_1=c_1+\frac{1}{N_c}c_2$, where $N_c=3$ is the number of colors. We take $\mu=m_b$ for $b$ decays and $a_1=1.14$, $a_2=-0.2$~\cite{PRD73-2006} are used in this work. To estimate the systematic uncertainties from non-factorizable contributions, we treat the $N_c$ as an adjustable parameter varying from 2 to to $+\infty$~\cite{wan-2015}, and then calculate the deviation to the central values. We stress that the factorization method used here is just taken as an preliminary study for the non-leptonic decays.

The annihilation matrix element can be expressed by decay constant and the momentum ($P_X$) or the polarization vector ($e^\mu$) of $X$ meson
\begin{subnumcases}
{\langle X|J^{\mu}|0\rangle=}
\si{i}f_P P^{\mu}_X &{X is a pseudoscalar meson},\\
f_V M_Xe^{\mu}      &{X is a vector~meson}.
\end{subnumcases}
$M_X$ is the mass of $X$ meson, $f_P$ and $f_V$ are the corresponding decay constants.

Finally, we get the non-leptonic decay width of the $B_c$ meson
\begin{equation} \label{eq-width2}
\Gamma=\frac{|\bm{p}|}{8\uppi M^2} \overline{|\mathcal{M}|^2},
\end{equation}
where $\bm{p}$ represents the 3-momentum of either of the two final mesons in the $B_c$ rest frame, which is expressed as $|\bm{p}|=\lambda^{\frac{1}{2}}(M^2, M^2_X, M^2_F)/(2M)$.

%******************** Section-3 *************************
\section{Hadronic Matrix Element}\label{Sec-3}
In this Section we will calculate the hadronic matrix element using the BS method. First we briefly review the instantaneous Bethe-Salpeter methods. Then we calculate the hadronic matrix transition element with the corresponding BS wave function. Finally the form factors are given graphically.
%******************3.1************
\subsection{Introduction to BS methods}
It is well known that the BS equation in momentum space reads~\cite{PR84-1951}
\begin{gather}\label{BS}
(\slashed p_1 - m_1)\Psi(q)(\slashed p_2+m_2)=\si{i}\int \frac{\text{d}^4k}{(2\uppi)^4}V(q-k)\Psi(k),
\end{gather}
where $\Psi(q)$ stands for the BS wave function; $V(q-k)$ is the BS interaction kernel; $p_1$ and $p_2$ are the momenta of constituent quark and anti-quark in the meson; $m_1$ and $m_2$ are the corresponding masses of constituent quark and anti-quark respectively~(see \autoref{Pic-1}). $p_1$ and $p_2$ can be described with the meson total momentum $P$ and inner relative momentum $q$ as
\begin{equation}
	\left \{
	\begin{aligned}
	p_1&=\alpha_1P+q,    &\qquad \alpha_1&=\frac{m_1}{m_1+m_2},\\
	p_2&=\alpha_2 P - q, &\qquad \alpha_2&=\frac{m_2}{m_1+m_2}.
	\end{aligned}
	\right .
\end{equation}

In the instantaneous approximation~\cite{PR87-1952}, $V(q-k)\sim V(|\bm q-\bm k|)$ does not depend on the time component of $(q-k)$. By using the same method in Ref.~\cite{PR87-1952}, we introduce the 3-dimensional Salpeter wave function $\varphi(q_\perp)$ and integration $\eta(q_\perp)$ as
\begin{gather}
\varphi(q_\perp)=\si{i}\int \frac{\text{d}q_P}{2\uppi}\Psi(q),\\
\eta(q_\perp)=\int \frac{\text{d}^3k_\perp}{(2\uppi)^3}V(|q_\perp-k_\perp|)\varphi(k_\perp),
\end{gather}
where $q_P=\frac{P\cdot q}{M}$ and $q_\perp=q-\frac{P}{M}q_P$, in rest frame of initial meson they correspond to the $q^0$ and $\bm q$ respectively; the integration $\eta(q_\perp)$ can be understood as the BS vertex for bound state. Now the BS equation (\ref{BS}) can be written as
\begin{gather}\label{BS-3D}
\Psi(q)=S(p_1)\eta(q_\perp)S(-p_2).
\end{gather}
$S(p_1)$ and $S(-p_2)$ are the propagators for the quark and anti-quark respectively, and can be decomposed as
\begin{equation}\label{Si}
\begin{aligned}
S(+p_1)&=\frac{\si{i}\Lambda_1^+}{q_P+\alpha_1M-\omega_1+\si{i}\epsilon}+\frac{\si{i}\Lambda_1^-}{q_P+\alpha_1M+\omega_1-\si{i}\epsilon},\\
S(-p_2)&=\frac{\si{i}\Lambda_2^+}{q_P-\alpha_2M+\omega_2-\si{i}\epsilon}+\frac{\si{i}\Lambda_2^-}{q_P+\alpha_2M-\omega_2+\si{i}\epsilon},
\end{aligned}
\end{equation}
where $\omega_i=\sqrt{m_i^2-q^2_\perp}~(i=1, 2)$ and projection operators $\Lambda^{\pm}_i(q_\perp)$~($i=1$ for quark and 2 for anti-quark) are defined as
\begin{gather}
\Lambda^{\pm}_i=\frac{1}{2\omega_i}\left[ \frac{\slashed P}{M}\omega_i\pm(-1)^{i+1}(m_i+\slashed q_\perp) \right].
\end{gather}

Since the BS kernel is instantaneous, we can perform contour integration over $q_P$ on both sides of Eq.~(\ref{BS-3D}) and then we obtain the coupled Salpeter equations~\cite{PR87-1952}
\begin{subnumcases}
~
(M-\omega_1-\omega_2)\varphi^{++}=+\Lambda_1^+(q_\perp)\eta(q_\perp)\Lambda^+_2(q_\perp),\label{BS-pp}\\
(M+\omega_1+\omega_2)\varphi^{--}=-\Lambda_1^-(q_\perp)\eta(q_\perp)\Lambda^-_2(q_\perp),\label{BS-nn}\\
\varphi^{+-}=\varphi^{-+}=0\label{BS-np}.
\end{subnumcases}
where $\varphi^{\pm\pm}$ are related to $\varphi$ by
\begin{gather}
\varphi^{\pm\pm}\equiv\Lambda_1^\pm(q_\perp)\frac{\slashed P}{M}\varphi(q_\perp)\frac{\slashed P}{M}\Lambda_2^\pm(q_\perp),\label{Def-np}\\
\varphi=\varphi^{++}+\varphi^{-+}+\varphi^{+-}+\varphi^{--}.\label{wave-sum}
\end{gather}
The normalization condition for BS equation now reads
\begin{gather}
\int \frac{\text{d}^3k_\perp}{(2\uppi)^3}\left[\overline\varphi^{++}\frac{\slashed P}{M}\varphi^{++}\frac{\slashed P}{M}-\overline\varphi^{--}\frac{\slashed P}{M}\varphi^{--}\frac{\slashed P}{M}\right]=2M.
\end{gather}

\subsection{Numerical results of Salpeter equations}
To solve the Salpeter equations numerically, first we choose the Cornell potential as the interaction kernel, which has the following forms~\cite{PLB584-2004}
\begin{equation}
\begin{aligned}
V(\bm q)&=(2\uppi)^3V_s(\bm q)+\gamma^0\otimes\gamma_0(2\uppi)^3V_v(\bm q),\\
V_s(\bm q)&=-(\frac{\lambda}{\alpha}+V_0)\delta^3(\bm q)+\frac{\lambda}{\uppi^2(\bm q^2+\alpha^2)^2},\\
V_v(\bm q)&=-\frac{2\alpha_s(\bm q)}{3\uppi^2(\bm q^2+\alpha^2)},\\
\alpha_s(\bm q)&=\frac{12\uppi}{27\ln(a+\frac{\bm q^2}{\Lambda_\text{QCD}})}.
\end{aligned}
\end{equation}
In above equations the symbol $\otimes$ denotes the BS wave function are sandwiched between the two $\gamma^0$ matrix. The model parameters we used are the same with before~\cite{JPG40-2013}, which reads
\begin{align*}
&a=e=2.7183,    	&\alpha&=0.06~\si{GeV}, & \lambda=0.21~\si{GeV}, \\
&m_c=1.62~\si{GeV},   &m_b&=4.96~\si{GeV},    &\Lambda_\text{QCD}=0.27~\si{GeV}.
\end{align*}

Now we just take the $0^{-}(^1\!S_0)$ state as an example to show how to solve the full coupled Salpeter equations to achieve the numerical results. The Salpeter wave function for $0^-(^1\!S_0)$ state has the following general form~\cite{PLB584-2004}
\begin{equation} \label{BS-wave-1s0}
\varphi(^1\!S_0)= M\bigg[ k_1 \frac{\slashed P}{M}+k_2 +k_3 \frac{\slashed q_{\perp}}{M}+k_4\frac{\slashed P \slashed q_{\perp}}{M^2}  \bigg ]\gamma^5.
\end{equation}
By utilizing the Salpeter equation~(\ref{BS-np}),  we can achieve the following two constraint conditions as
\begin{equation}\label{BS-par-1s0}
\begin{aligned}
k_3=& +\frac{M(\omega_1-\omega_2)}{m_1\omega_2+m_2\omega_1}k_2,\\
k_4=& -\frac{M(\omega_1+\omega_2)}{m_1\omega_2+m_2\omega_1}k_1.
\end{aligned}
\end{equation}
Now in above $^1\!S_0$ state Salpeter wave function, there are only two undetermined wave function $k_1$ and $k_2$, which are just the functions of $q^2_\perp$.

By using the definition Eq.~(\ref{Def-np}), the positive wave function for the ${^1\!S_0}$ state can be written as
\begin{equation} \label{wave-1s0}
\varphi^{++}(^1\!S_0)=\bigg [ A_1+A_2 \frac{\slashed P}{M} +A_3 \frac{\slashed q_{\perp}}{M}+A_4\frac{\slashed P \slashed q_{\perp}}{M^2}  \bigg ]\gamma^5.
\end{equation}
$A_i~(i=1,2,3,4)$ have the following forms
\begin{equation}\label{par-1s0}
\begin{aligned}
A_1=& \frac{M}{2}\biggl[\frac{\omega_1+\omega_2}{m_1+m_2}k_1+k_2 \bigg],\\
A_2=&\frac{M}{2}\biggl[k_1+\frac{m_1+m_2}{\omega_1+\omega_2}k_2 \bigg],\\
A_3=&-\frac{M(\omega_1-\omega_2)}{m_1\omega_2+m_2\omega_1}A_1,\\
A_4=&-\frac{M(m_1+m_2)}{m_1\omega_2+m_2\omega_1}A_1.
\end{aligned}
\end{equation}
Similarly, the $\varphi^{--}(^1\!S_0)$ is expressed as
\begin{equation} \label{wave-1s0-nn}
\varphi^{--}(^1\!S_0)=\bigg [ Z_1+Z_2 \frac{\slashed P}{M} +Z_3 \frac{\slashed q_{\perp}}{M}+Z_4\frac{\slashed P \slashed q_{\perp}}{M^2}  \bigg ]\gamma^5.
\end{equation}
$Z_i~(i=1,2,3,4)$ has the following forms
\begin{equation}\label{par-1s0-nn}
\begin{aligned}
Z_1=& \frac{M}{2}\biggl[k_2 -\frac{\omega_1+\omega_2}{m_1+m_2}k_1\bigg]    ,\\
Z_2=&\frac{M}{2}\biggl[k_1-\frac{m_1+m_2}{\omega_1+\omega_2}k_2 \bigg]    ,\\
Z_3=&-\frac{M(\omega_1-\omega_2)}{m_1\omega_2+m_2\omega_1}Z_1,\\
Z_4=&+\frac{M(m_1+m_2)}{m_1\omega_2+m_2\omega_1}Z_1.
\end{aligned}
\end{equation}
And now the normalization condition reads
\begin{gather}\label{Norm-1S0}
\int \frac{\text{d}^3\bm{q}}{(2\uppi)^3}\frac{8M\omega_1\omega_2k_1k_2}{(m_1\omega_2+m_2\omega_1)}=1.
\end{gather}
Inserting the expressions of $\varphi^{++}(^1\!S_0)$ and $\varphi^{--}(^1\!S_0)$ into Eq.~(\ref{BS-pp}) and~(\ref{BS-nn}), respectively, we can obtain the two coupled eigen equations on $k_1$ and $k_2$~\cite{PLB584-2004} as
\begin{equation}
\left \{
\begin{aligned}
(M-\omega_1-\omega_2)\left[ck_1(\bm q)+k_2(\bm q)\right]=\frac{1}{2\omega_1\omega_2}\int \text{d}^3\bm{k} \left[H_{1}k_1(\bm k)+H_{2}k_2(\bm k)\right],\\
(M+\omega_1+\omega_2)\left[k_2(\bm q)-ck_1(\bm q)\right]=\frac{1}{2\omega_1\omega_2}\int \text{d}^3\bm{k} \left[H_{1}k_1(\bm k)-H_{2}k_2(\bm k)\right],
\end{aligned}
\right .
\end{equation}
where we have used definition $c=\frac{\omega_1+\omega_2}{m_1+m_2}$ and the shorthand
\begin{equation}
\begin{aligned}
H_1&=\bm k\cdot \bm q(V_s+V_v)\frac{(\nu_1+\nu_2)(\omega_1+\omega_2)}{m_1\nu_2+m_2\nu_1} -(V_s-V_v)(m_1\omega_2+m_2\omega_1),\\
H_2&=\bm k\cdot \bm q(V_s+V_v)\frac{(\nu_1-\nu_2)(m_1-m_2)}{m_1\nu_2+m_2\nu_1}           -(V_s-V_v)(m_1m_2+\omega_1\omega_2+\bm q^2).
\end{aligned}
\end{equation}
In above equations we have defined $\nu_i=\sqrt{m^2_i+\bm k^2}~(i=1,2)$.
Then by solving the two coupled eigen equations, we achieve the mass spectrum and corresponding wave functions $k_1$ and $k_2$. Repeating the similar procedures we can obtain the numerical wave functions for $2^{-+}(^1\!D_2)$, $2^{--}(^3\!D_2)$ and $3^{--}(^3\!D_3)$. Interested reader can see more details on solving the full Salpeter equations in Refs.~\cite{3D23-wave,JPG40-2013,PLB584-2004}.

%***********************************************************************
% sub-section 2
\subsection{Form factors for hadronic transition}
Now we will calculate the form factors with BS methods. According to Mandelstam formalism~\cite{PRSA233-1955}, the hadronic transition matrix element $\langle c\bar c |h^{\mu}|B_c \rangle$ can be directly written as
\begin{equation} \label{A-h}
\begin{aligned}
\langle c\bar c |h^{\mu}|B_c \rangle=&\mathrm{i} \int \frac{\mathrm{d}^4 q \mathrm{d}^4 q'}{(2 \uppi)^4} \mathrm{Tr} [\bar{\Psi}(q') \Gamma^{\mu} \Psi(q) S^{-1}(-p_2) \delta^{(4)}(p_2-p'_2)] \\
            =&\mathrm{i} \int \frac{\mathrm{d}^4 q}{(2 \uppi)^4} \mathrm{Tr} [\bar{\Psi}(q') \Gamma^{\mu} \Psi(q) S^{-1}(-p_2)].
\end{aligned}
\end{equation}
In the above expression, $\Psi(q')$ stands for the BS wave function of final $c\bar c$ systems and $\bar \Psi=\gamma^0\Psi^\dagger\gamma^0$; $q'$ is the inner relative momentum of $c \bar c$ system, which is related to the quark (anti-quark) momentum $p_1^{\prime}$ ($p_2^{\prime}$) by $p'_i=\alpha^\prime_i P_F+(-1)^{i+1}q^\prime$ and $\alpha^\prime_i=\frac{m^\prime_i}{m'_1+m'_2}$~($i=1, 2$), where $m'_i$ are masses of the constituent quarks in the final bound states~(see \autoref{Pic-1}); here we have $m_1=m_b$, $m_2=m'_2=m'_1=m_c$; $S^{-1}(-p_2)=(-\slashed p_2-m_2)$ is the inverse of propagator for anti-quark. Since the the propagator $S_2$ is used by both initial and final mesons, here we add an $S^{-1}(-p_2)$ factor.  As there is a delta function in the first line of the above equation, the relative momenta $q$ and $q^\prime$ are related by $q '=q-(\alpha_2 P-\alpha '_2 P_F)$.

By inserting Eq.~(\ref{BS-3D}) and (\ref{Si}) into Eq.~(\ref{A-h}), then perform the counter integral over $q_P$ and we get
\begin{align}
\langle c\bar c |h^{\mu}|B_c \rangle=&\int \frac{d^3 q_\perp}{(2\uppi)^3} \text{Tr}\bigg\lbrace \frac{\slashed{P}}{M} (   \bar{\varphi}'^{++} \Gamma^\mu \varphi^{++} + \bar{\varphi}'^{++} \Gamma^\mu \psi^{-+} - \bar{\psi}'^{-+} \Gamma^\mu \varphi^{--}+ \bar{\psi}'^{+-}\Gamma^\mu \varphi^{++}  \\
 &\qquad \qquad \qquad
 - \bar{\varphi}'^{--} \Gamma^\mu \psi^{+-} - \bar{\varphi}'^{--} \Gamma^\mu \varphi^{--})  \bigg\rbrace,
\end{align}
where we have used the following definitions
\begin{equation}
    \begin{aligned}
% \varphi^{++}   &= \frac{\Lambda^+_1       \eta  \Lambda^+_2}{(M- \omega_1 - \omega_2)}, 		  &\quad  \bar{\varphi'}^{++}  &= \frac{\Lambda'^+_2 \bar{\eta}' \Lambda'^+_1}{E'- \omega'_1 - \omega'_2},   \\
% \varphi ^{--}  &= \frac{-\Lambda^-_1      \eta  \Lambda^-_2}{(M+ \omega_1 + \omega_2)},  		  &\quad  \bar{\varphi'}^{--}  &= \frac{-\Lambda'^-_2 \bar{\eta}' \Lambda'^-_1}{(E'+ \omega'_1 + \omega'_2)},\\
\psi^{-+}   	&= \frac{\Lambda^-_1        \eta   \Lambda^+_2}{(\omega'_1 + \omega_1)+ (M-E')}, &\quad  \bar{\psi}'^{-+}    &= \frac{\Lambda'^-_2 \bar{\eta}' \Lambda'^+_1}{(\omega'_1 + \omega_1)+ (M-E')},\\
\psi^{+-}   	&= \frac{\Lambda^+_1        \eta   \Lambda^-_2}{(\omega'_1 + \omega_1) - (M-E')}, &\quad  \bar{\psi}'^{+-}    &= \frac{\Lambda'^+_2 \bar{\eta}' \Lambda'^-_1}{(\omega'_1 + \omega_1) - (M-E')}.
    \end{aligned}
\end{equation}
$\varphi^{++}$ is the Salpeter positive wave function, which is much larger than $\psi^{-+},\psi^{+-}$ and $\varphi^{--}$ in the case of weak binding~\cite{cch1,Durand-1982}. In the following calculations we will only consider the dominant $[\bar \varphi'^{++}\Gamma^\mu\varphi^{++}]$ part, while others' contributions are ignored. The reliability of this approximation can be seen in Ref.~\cite{NPP39-2012}.
Finally we obtain the form factors described with 3-dimensional Salpeter positive wave function
\begin{equation} \label{Ah}
\langle c\bar c |h_{bc}^{\mu}|B_c \rangle=\int \frac{d^3 q_\perp}{(2\uppi)^3} \text{Tr} \bigg [\frac{\slashed P}{M} \bar \varphi'^{++}(q'_\perp) \Gamma^\mu \varphi^{++}(q_\perp)  \bigg ],
\end{equation}

In our calculation, the final charmonium states are ${^1\!D_2}(2^{-+}),~ {^3\!D_2}(2^{--})$, or ${^3\!D_3}(3^{--})$. Their BS wave functions are constructed by considering the spin and parity of the corresponding mesons~\cite{BS-wave}. We will take the ${^1\!D_2} (2^{-+})$ state as an example to show how to do the calculation to achieve form factors. The results of other mesons will be given directly.

The Salpeter wave function of ${^1\!D_2}$ states with equal mass can be written as \cite{JPG40-2013}
\begin{equation}\label{wave-f1d2}
\varphi_{2^{-+}}=e^{\mu \nu}q'_{\mu\perp}q'_{\nu\perp}\bigg[f_1+f_2 \frac{{\slashed P}_{\!F}}{M_F}+f_4 \frac{\slashed P_F \slashed q'_\perp}{M_F^2}\bigg].
\end{equation}
And Salpeter equation~(\ref{BS-np}) gives the constraint condition $f_4= -\frac{M_F}{m_c}f_2$, where $m_c$ is the $c$ quark constituent mass; $e^{\mu \nu}$ is the symmetric polarization tensor for $J=2$, which satisfies the following relations~\cite{PRD43-1991}
\begin{equation}
e^{\mu \nu}  {P_F}_{\mu}=0,\quad e^{\mu \nu} g_{\mu \nu}=0.
\end{equation}
And the completeness relation for the polarization tensor is
\begin{equation}\label{Polar2}
\sum^2_{m=-2} e^{\mu \nu}(m)e^{\alpha \beta}(m)=\frac{1}{2} \big (g_{\perp}^{\alpha \mu}g_{\perp}^{\beta \nu}+g_{\perp}^{\alpha \nu}g_{\perp}^{\beta \mu}\big )-\frac{1}{3}g_{\perp}^{\alpha \beta}g_{\perp}^{\mu \nu},
\end{equation}
where we have defined $g_{\perp}^{\alpha \beta}\equiv-g^{\alpha \beta}+\frac{P_F^\alpha P_F^\beta}{P_F^2}$.

From the definition, we get the Salpeter positive wave function for ${^1\!D_2}(2^{-+})$ charmonium~\cite{JPG40-2013} as
\begin{equation} \label{wave-1d2}
\varphi^{++}(^1\!D_2)=e^{\mu \nu} q'_{\mu\perp} q'_{\nu\perp} \bigg [ B_1+B_2 \frac{\slashed P_F}{M_F} +B_4\frac{\slashed P_F \slashed q'_{\perp}}{M_F^2} \bigg]\gamma^5.
\end{equation}
%$B_i~(i=1,2,4)$ have the following expressions
\begin{equation}\label{par-1d2}
\begin{aligned}
B_1=& \frac{1}{2}\Big[f_1+\frac{\omega_c}{m_c}f_2  \Big]    ,\\
B_2=& \frac{1}{2}\Big[f_2+\frac{m_c}{\omega_c}f_1  \Big]    ,\\
B_4=& -\frac{M_F}{\omega_c}B_1,
\end{aligned}
\end{equation}
where $\omega_c=\sqrt{m^2_c-q'^2_\perp}$; $f_1$ and $f_2$ are functions of $q'_\perp$.

Having theses wave functions, we can deal with the form factors in the hadronic matrix element. For the transition $B_c\To\eta_{c2}$, inserting Eq.~(\ref{wave-1s0}) and Eq.~(\ref{wave-1d2}) into Eq.~(\ref{Ah}) and finishing the trace, we obtain the form factors $s_1,s_2, s_3$ and $s_4$ in Eq.~(\ref{form-1})
\begin{equation}\label{si-1}
\begin{aligned}
s_1 &=\int \frac{d^3 \bm q}{(2\uppi)^3} \bigg[x_1-\frac{{C_1} E_F ({x_3}+{x_4})}{M p_F}+\frac{({x_6}+{x_7}) ({C_{21}} E_F^2-{C_{22}} p_F^2)}{M^2 p_F^2}+\frac{E_F {x_9} (3 {C_{32}} p_F^2-{C_{31}} E_F^2)}{M^3 p_F^3}\bigg],\\
s_2 &=\int \frac{d^3 \bm q}{(2\uppi)^3} \bigg[x_2+\frac{{C_1} (M {x_3}-E_F {x_5})}{M p_F}+\frac{{C_{21}} E_F (E_F {x_8}-M {x_6})}{M^2 p_F^2}-\frac{{C_{22}} {x_8}}{M^2}+\frac{{x_9}({C_{31}} E_F^2-C_{32}p^2_F)}{M^2 p_F^3}\bigg],\\
s_3 &=\int \frac{d^3 \bm q}{(2\uppi)^3} ({C_{22}} {x_6}-\frac{2 {C_{32}} E_F {x_9}}{M p_F}) ,\\
s_4 &=\int \frac{d^3 \bm q}{(2\uppi)^3} ({C_{22}} {x_{10}}-\frac{2 {C_{32}} E_F {x_{11}}}{M p_F}).
\end{aligned}
\end{equation}
In the above expressions, $p_F$ denotes the absolute value of $\bm{P}_F$ which is the 3-momentum of the final charmonium, $E_F=\sqrt{M_F^2+p_F^2}$. The specific expressions of $x_i~(i=1,2,\cdots,11)$ can be found in~\ref{xi}. $C_i$ are expressed as
\begin{equation}\label{eq-Ci}
\left \{
\begin{aligned}
C_{1~}=&~ |\bm q| \cos \eta,                                  &\qquad            C_{21}=&~\frac{1}{2}|\bm q|^2(3\cos^2 \eta-1),\\
C_{22}=&~\frac{1}{2}|\bm q|^2(\cos^2 \eta-1),                 &\qquad            C_{31}=&~\frac{1}{2}|\bm q|^3(5\cos^3 \eta-3 \cos \eta),\\
C_{32}=&~\frac{1}{2}|\bm q|^3(\cos^3 \eta- \cos \eta),      &\qquad            C_{41}=&~\frac{1}{8}|\bm q|^4(35\cos^4 \eta-30\cos^2 \eta+3),\\
C_{42}=&~\frac{1}{8}|\bm q|^4(5\cos^4 \eta-6\cos^2 \eta+1), &\qquad            C_{43}=&~\frac{1}{8}|\bm q|^4(\cos^4 \eta-2\cos^2 \eta+1),
\end{aligned}
\right .
\end{equation}
where $\eta$ is the angle between $\bm{q}$ and $\bm{P}_F$.

\begin{figure}[ht]
\centering
\subfigure[Form factors of $B_c\To \eta_{c2}(1^1\!D_2)$.]{\includegraphics[width=0.48\textwidth]{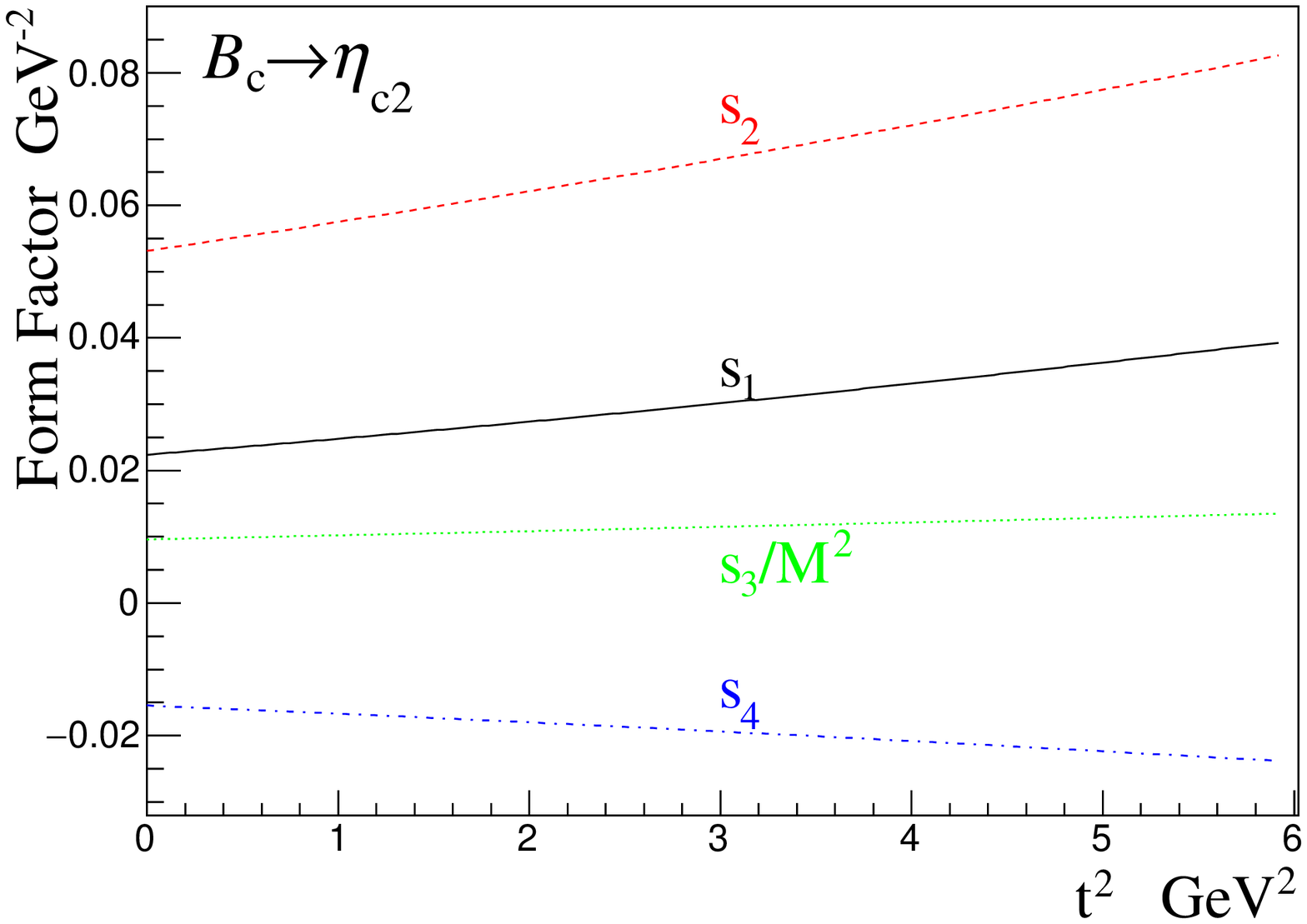}\label{Fig-etac2}}
\subfigure[Form factors of $B_c\To \psi_{2}(1^3\!D_2)$.]{\includegraphics[width=0.48\textwidth]{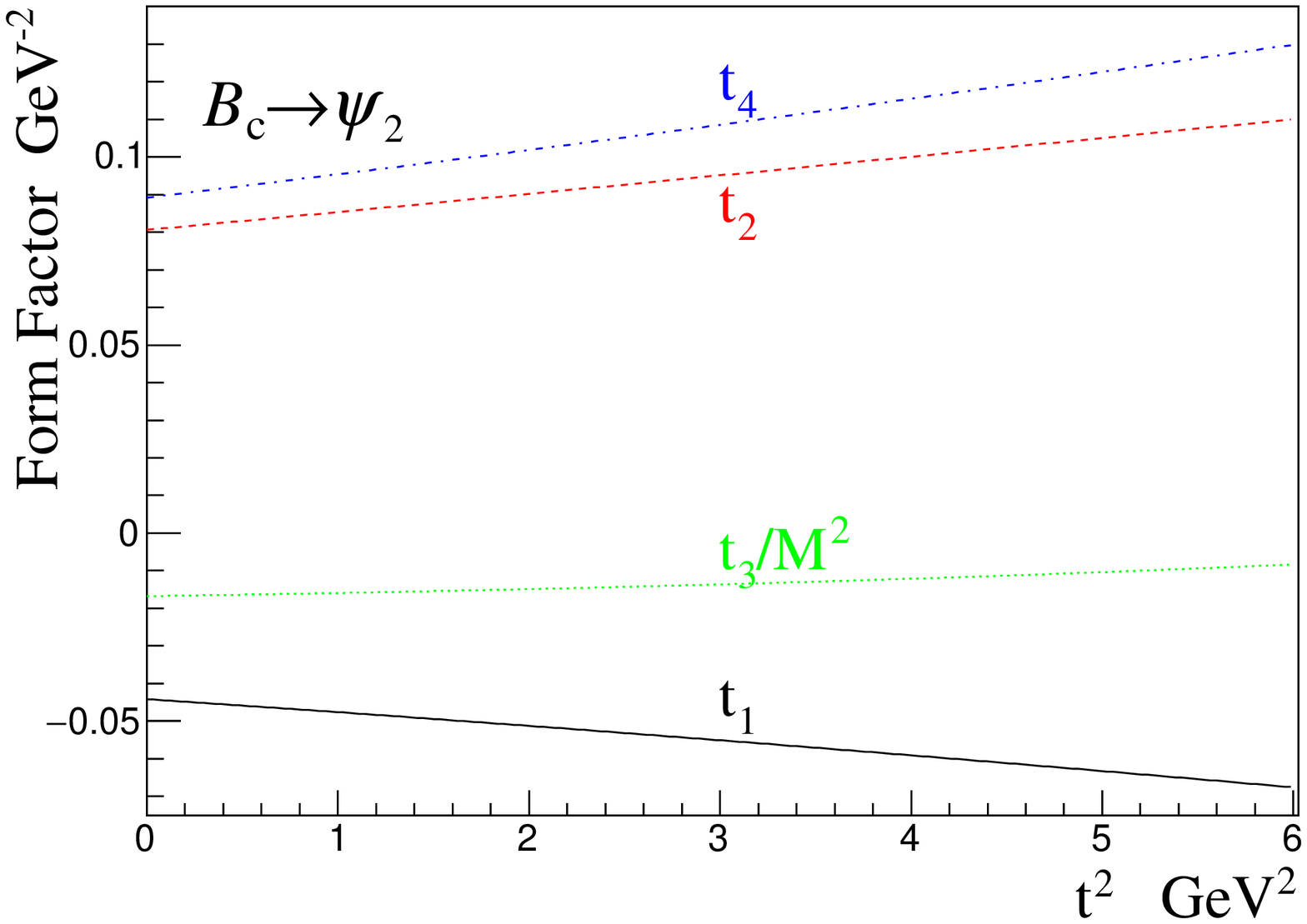}\label{Fig-psi2}}
\subfigure[Form factors of $B_c\To \psi_{3}(1^3\!D_3)$.]{\includegraphics[width=0.48\textwidth]{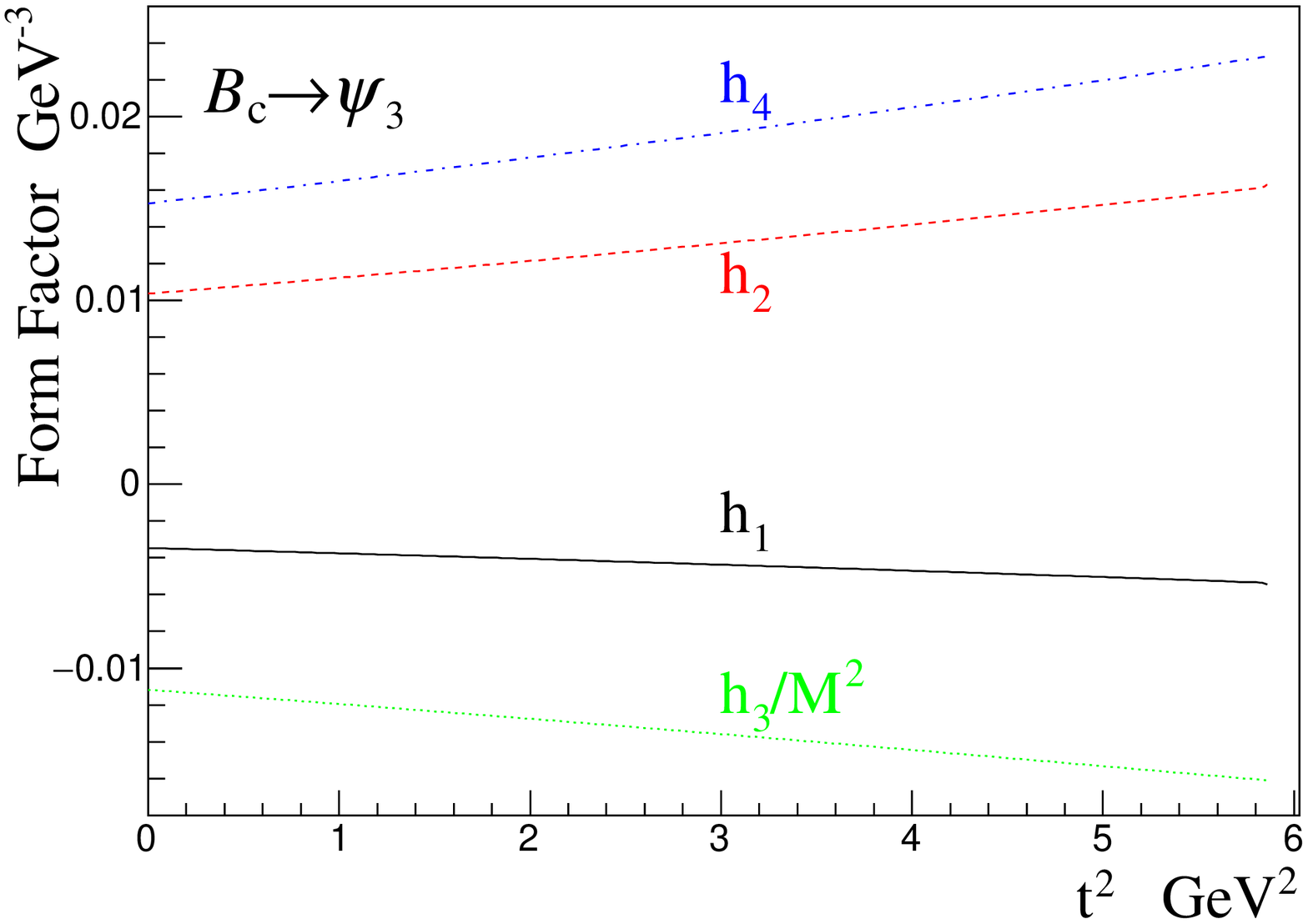}\label{Fig-psi3}}
\caption{Form factors for $B_c\To \eta_{c2}, \psi_2$ and $\psi_3$. $t^2=(P-P_F)^2$ and $t$ denotes the transferred momentum. We have divided $s_3, t_3$ and $h_3$ by $M^2$ to keep their dimensions consistent with others'.}\label{Fig-form}
\end{figure}

Replacing the wave function $\varphi^{++}(^1\!D_2)$ by $\varphi^{++}(^3\!D_2)$ or $\varphi^{++}(^3\!D_3)$, and repeating the procedures above, we can get the form factors for the transition of $B_c$ to $\psi_2(1^3\!D_2)$ or $\psi_3(1^3\!D_3)$ charmonium. The Salpeter positive wave function for $2^{--}(^3\!D_2)$ and $3^{--}(^3\!D_3)$~\cite{3D23-wave} can be seen in~\ref{fun-3d2-3d3}. We will not give the bulky analytical expressions but only present the form factors for the decays to $^3\!D_2$ and $^3\!D_3$ charmonia graphically~(see~\autoref{Fig-form}).

Finally we can obtain the numerical results of form factors. In \autoref{Fig-etac2}$\sim$ \autoref{Fig-psi3}, we show the form factors $s_i$, $t_i$ and $h_i$ $(i=1,2,3, 4)$ which change with momentum transfer $t^2$, where $t^2=(P-P_F)^2$. To make the form factors have the same dimension, we have divided $s_3$, $t_3$ and $h_3$ by $M^2_{B_c}$.
%One can see that for the $J=2$ case, compared to the other three form factors, $s_3$ and $t_3$ change very slow as $(P-P_F)^2$ increases. For $\eta_{c2}$, $s_1$ and $s_2$ are increasing functions, while $s_4$ is a decreasing function. For $\psi_2$, $t_1$ ($t_2$ and $t_4$) is decreasing (increasing) function. For the $J=3$ case, namely, $\psi_3$, $h_1$ and $h_3$ ($h_2$ and $h_4$) are decreasing (increasing) functions. we can see that except $\eta_{c2}$, for which only one form factor is negative, other two cases have two negative form factors, respectively.
One can notice that the form factors we got are quite smooth in all the concerned range of $t^2$. This is important for the calculation of non-leptonic decays, which depends sensitively on one specific point of the form factors.

\section{Decay width and Discussions}\label{Sec-4}
For the $\psi_2(1^3\!D_2)$ meson, which has been found experimentally to be $X(3823)$~\cite{Belle-2013}. For $\eta_{c2}(1^1\!D_2)$ and $\psi_3(1^3\!D_3)$, we use the predictions of Ref.~\cite{PRD72-2005}.
The meson masses we used in this work are
\begin{gather*}
M_{B_c}= 6.276~\si{GeV},\quad M_{\eta_{c2}}=3.837~\si{GeV}, \quad M_{\psi_2}= 3.823~\si{GeV}, \quad
M_{\psi_3} = 3.849~\si{GeV}.
\end{gather*}
The lifetime for $B_c$ meson is $\tau_{_{Bc}}=0.452\times 10^{-12}$~\si{s}~\cite{PDG-2014}. The values of CKM matrix elements we use in this work are
\begin{align*}
 V_{cb}=0.041, \quad V_{ud}=0.974, \quad V_{us}=0.225.
\end{align*}

Among the three $D$-wave charmonia we calculated here, $\psi_2(1^3\!D_2)$ and $\eta_{c2}(1^1\!D_2)$ are expected to be quite narrow since there are no open charm decay modes. Both of them are just above the threshold of $D\bar{D}$ while below $D\bar{D}^*$. However, the conservation of parity forbids the $D\bar{D}$ channel. So the dominant decay modes are expected to be electromagnetic ones. For $\psi_2(1^3\!D_2)$, the total width are estimated to be $\sim0.4$~\si{MeV}~\cite{PRD55-1997}. The predominant EM decay channel of this particle is $\eta_{c2}(1^1\!D_2)\!\to \! h_c(1P)\gamma$ and the corresponding decay width is about $0.3$~\si{MeV}~\cite{PRD67-2003,PRD79-2009}. For $\psi_3(1^3\!D_3)$,  although its mass is above the $D\bar{D}$ threshold, the decay width is estimated to be less than 1~\si{MeV}~\cite{PRD72-2005, PRD73-014014-2006}. The reasons are that the phase space is small and there is a $F$-wave centrifugal barrier. The radiative width for the main EM transition $\psi_3(1^3\!D_3)\To\gamma\chi_{c2}$ is $\sim0.3$~\si{MeV}.

\begin{table}[ht]
\caption{Branching ratios of $B_c$ semi-leptonic decays. The uncertainties here are determined by varying the model parameters by $\pm5\%$ and then finding the maximum deviation.}\label{semiB}
\vspace{0.2em}\centering
\begin{tabular}{ccccc}
%&&&&\hfill{($10^{-4}$)}\\
\toprule[1.5pt]
       Channels                  & Ours     & Ref.~\cite{PRD71-2005} & Ref.~\cite{PRD73-2006}& Ref.~\cite{PRD74-2006}   \\
\midrule[1.2pt]  \vspace{0.4em}
$B_c^-\To \eta_{c2}  e  \bar{\nu} $    & $5.9^{-0.8}_{+1.0}\e{-4}$   &		-		&           -      &      -           \\  \vspace{0.3em}
$B_c^-\To \eta_{c2} \mu  \bar{\nu}$    & $5.8^{-0.8}_{+1.0}\e{-4}$   &		-		&           -      &      -          \\ \vspace{0.3em}
$B_c^-\To \eta_{c2} \tau \bar{\nu}$    & $4.9^{-0.8}_{+1.0}\e{-6}$   &		-		&           -      &      -          \\  \vspace{0.3em}
$B_c^-\To \psi_2    e\bar{\nu}    $    & $1.5^{-0.2}_{+0.3}\e{-4}$   & $8.9\e{-5}$		& $6.6\e{-5}$      &     $4.3^{-0.5}\e{-5}$\\ \vspace{0.3em}
$B_c^-\To \psi_2    \mu \bar{\nu} $    & $1.5^{-0.2}_{+0.3}\e{-4}$   &		-		&           -      &      -           \\ \vspace{0.3em}
$B_c^-\To \psi_2    \tau \bar{\nu}$    & $2.3^{-0.4}_{+0.5}\e{-6}$   &  $2.1\e{-6}$		& $9.9\e{-7}$      &     $8.3^{-1.0}\e{-7}$       \\ \vspace{0.3em}
$B_c^-\To \psi_3    e    \bar{\nu}$    & $3.5^{-0.6}_{+0.8}\e{-4}$   &		-		&           -      &      -           \\  \vspace{0.3em}
$B_c^-\To \psi_3    \mu \bar{\nu} $    & $3.4^{-0.6}_{+0.7}\e{-4}$   &		-		&           -      &      -           \\  \vspace{0.1em}
$B_c^-\To \psi_3    \tau \bar{\nu}$    & $2.3^{-0.5}_{+0.6}\e{-6}$   &		-		&           -      &      -           \\
\bottomrule[1.5pt]
\end{tabular}
\end{table}
%\vspace{1ex}

\subsection{Branching ratios and lepton spectra for $B_c$ semi-leptonic decays}
From the results of form factors, we can get the branching ratios of $B_c$ exclusive decays. The semi-leptonic decay widths of $B_c$ to $D$-wave charmonia are list in \autoref{semiB}.
For the theoretical uncertainties, here we will just discuss the dependence of the final results on our model parameters $\lambda,~\Lambda_\text{QCD},~m_b$ and $m_c$ in the Cornell potential. The theoretical errors, induced by these four parameters, are determined by varying every parameter by $\pm5\%$, and then scanning the four-parameter space to find the maximum deviation. Generally, this theoretical uncertainties can amount to $10\%\sim20\%$ for $B_c$ semi-leptonic decays.

Our result for the branching ratio of the channel $B_c\To \psi_{2}e\bar{\nu}_e$ is $1.5\e{-4}$, which is larger than those of Refs.~\cite{PRD71-2005, PRD73-2006} and Ref.~\cite{PRD74-2006}. For the channel with $\tau$ as the final lepton, our result is very close to that in Ref.~\cite{PRD71-2005}, but more than two times larger than those of Refs.~\cite{PRD73-2006, PRD74-2006}. The method used in Ref.~\cite{PRD74-2006} is non-relativistic constituent quark model. Both Ref.~\cite{PRD71-2005} and Ref.~\cite{PRD73-2006} used the same relativistic constituent quark model %(but with different parameters which leads to the different results in \autoref{semiB})
whose framework is relativistic covariant while the wave functions of mesons are assumed to be the Gaussian type. As to our method, although the instantaneous approximation causes the lost of relativistic covariant, the wave functions are more reasonable.  For the $\eta_{c2}$ and $\psi_3$ cases, we get $\mathcal{B}(B_c\To \eta_{c2}e\bar{\nu}_e)=5.9\times 10^{-4}$ and $\mathcal{B}(B_c\To\psi_{3}e\bar{\nu}_e)=3.5\times 10^{-4}$ which are larger than that of the $\psi_2$ case. From this point, the former two channels have more possibilities to be detected in the future experiments.

\begin{table}[ht]
\caption{$A_{FB}$ of $B_c$ semi-leptonic decays.}\label{Tab-AFB}
\vspace{0.2em}\centering
\begin{tabular}{lrcc}
\toprule[1.5pt]
       Channels                  & Ours     & Ref.~\cite{PRD71-2005}& Ref.~\cite{PRD74-2006}   \\
\midrule[1.2pt]
$B_c^-\to \eta_{c2}  e  \bar{\nu} $    & -0.020   &           -      &      -                      \\
$B_c^-\to \eta_{c2} \mu  \bar{\nu}$    & 0.011  	 &           -      &      -                      \\
$B_c^-\to \eta_{c2} \tau \bar{\nu}$    & 0.35    &           -      &      -                      \\
$B_c^-\to \psi_2    e\bar{\nu}    $    & -0.56   &          -0.21   &     -0.59                    \\
$B_c^-\to \psi_2    \mu \bar{\nu} $    & -0.56   &           -      &     -0.59                      \\
$B_c^-\to \psi_2    \tau \bar{\nu}$    & -0.37	 &          -0.21   &     -0.42                  \\
$B_c^-\to \psi_3    e    \bar{\nu}$    & -0.11   &           -      &      -                      \\
$B_c^-\to \psi_3    \mu \bar{\nu} $    & -0.090   &           -      &      -                      \\
$B_c^-\to \psi_3    \tau \bar{\nu}$    & 0.10    &           -      &      -                      \\
\bottomrule[1.5pt]
\end{tabular}
\end{table}

As an experimentally interested quantity, the numerical results for the forward-backward asymmetry $A_{FB}$ are list in \autoref{Tab-AFB}. For the $B_c\To\psi_2 \ell{\bar{\nu}}$ channel, our results are consistent with those in Ref.~\cite{PRD74-2006} but larger than those in Ref.~\cite{PRD71-2005}. We notice that for all the cases when $\ell=e$, $\mu$, and $\tau$, $A_{FB}(\psi_2)$ is negative. For the $B_c\To\eta_{c2} \ell\bar{\nu}$ channel, when $\ell=e$, $A_{FB}(\eta_{c2})$ is negative, while for the $B_c\To\psi_3 \ell\bar{\nu}$ channel, when $\ell=e$ and $\mu$, $A_{FB}(\psi_3)$ is negative. For the absolute value of this quantity, when $\ell=e$, we have $A_{FB}(\eta_{c2})<A_{FB}(\psi_3)<A_{FB}(\psi_2)$.

\begin{figure}[ht]
\subfigure[Angular spectrum for decay to $e$ mode.]      {\includegraphics[width=0.48\textwidth]{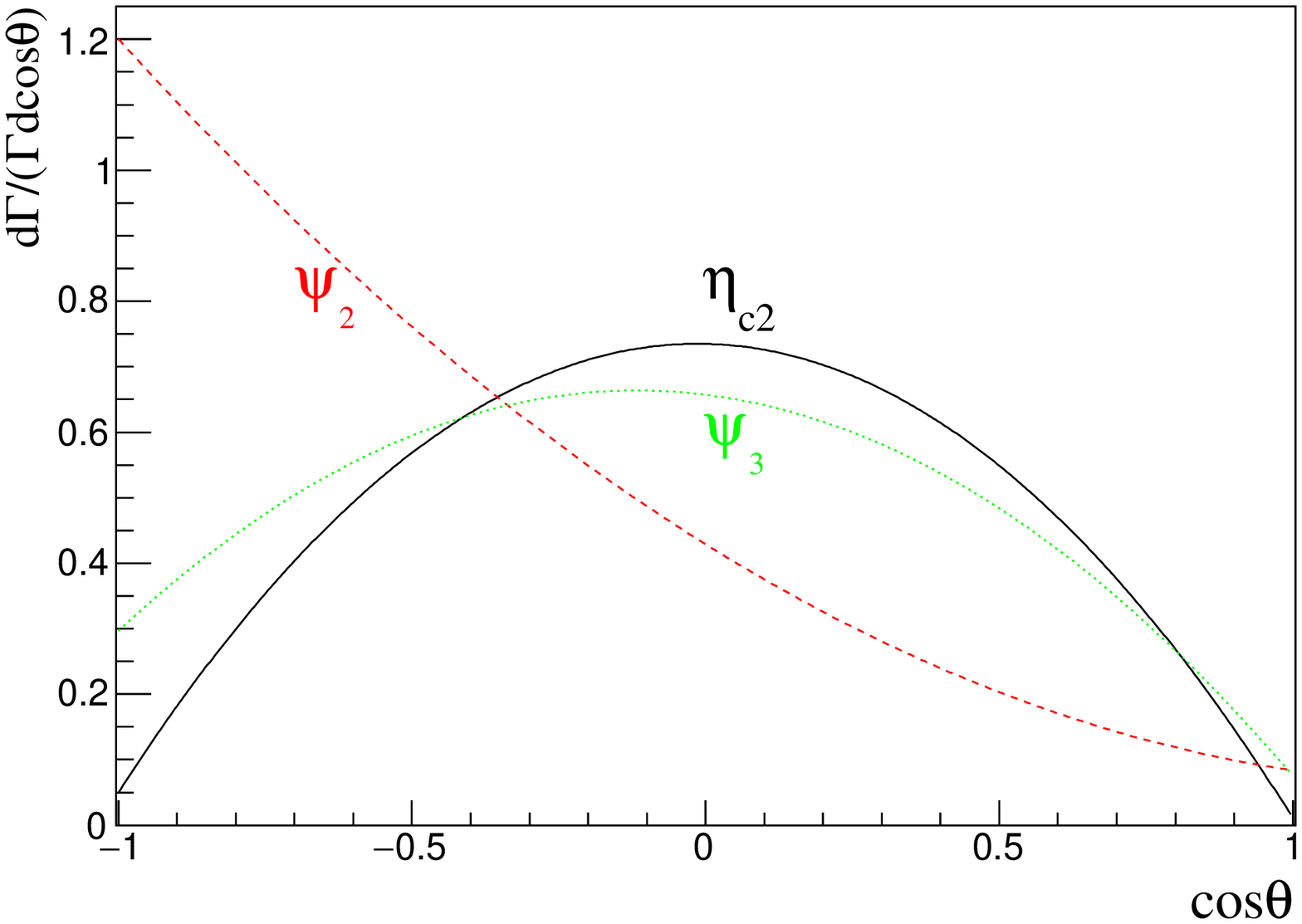} \label{Fig-dwe-cos}}
\subfigure[Angular spectrum for decay to $\tau$ mode.]   {\includegraphics[width=0.48\textwidth]{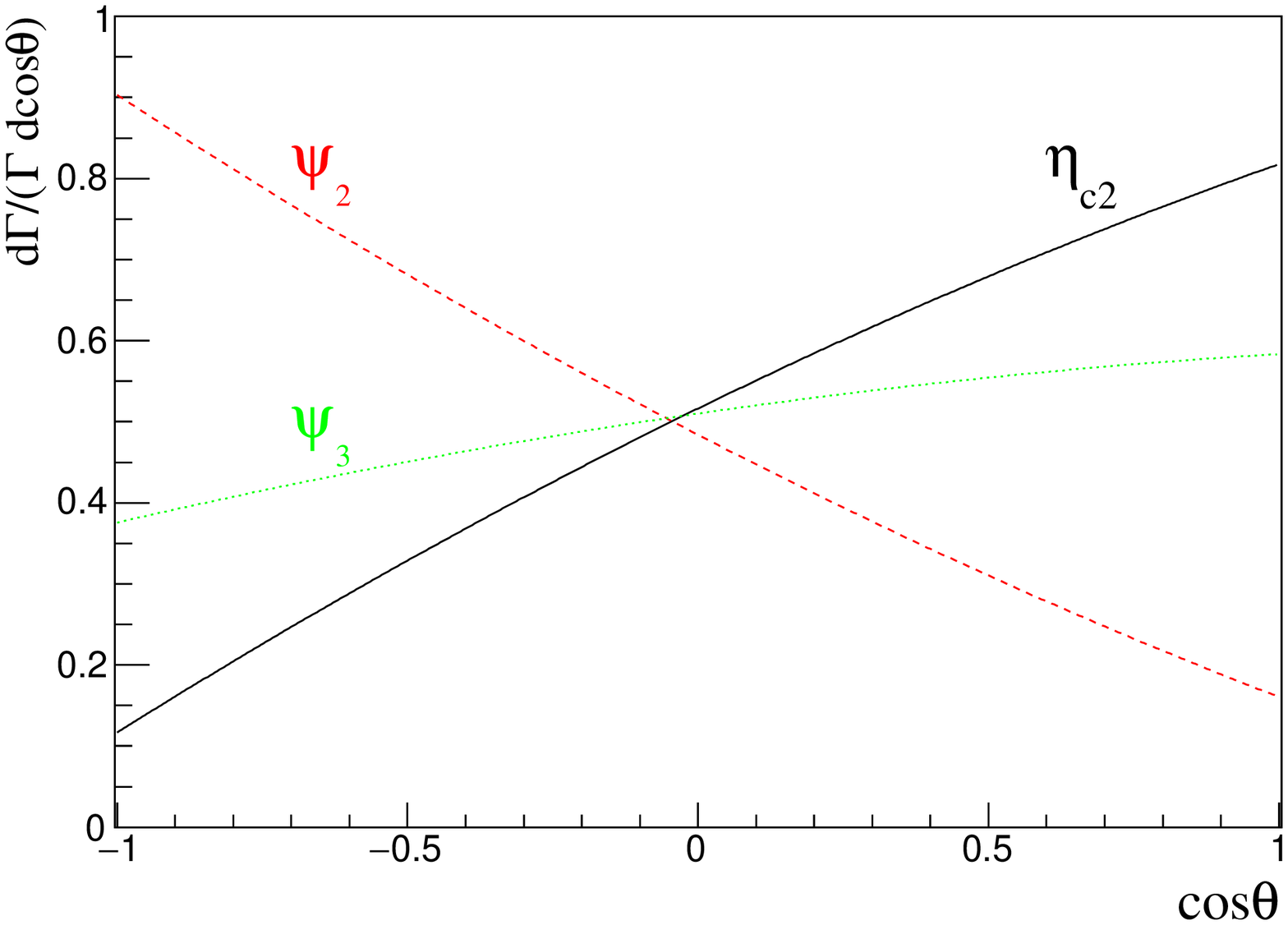} \label{Fig_dwt-cos}}
\caption{The spectra of relative width vs $\cos\theta$ in $B_c$ semi-leptonic decays into $D$-wave charmonia. $\theta$ is the angle between charged lepton $\ell$ and final $c\bar c$ system in the rest frame of $\ell\bar{\nu}$.} \label{Fig-cos}
\end{figure}

\begin{figure}[ht]
\subfigure[Momentum spectrum for decay to $e$ mode.]      {\includegraphics[width=0.48\textwidth]{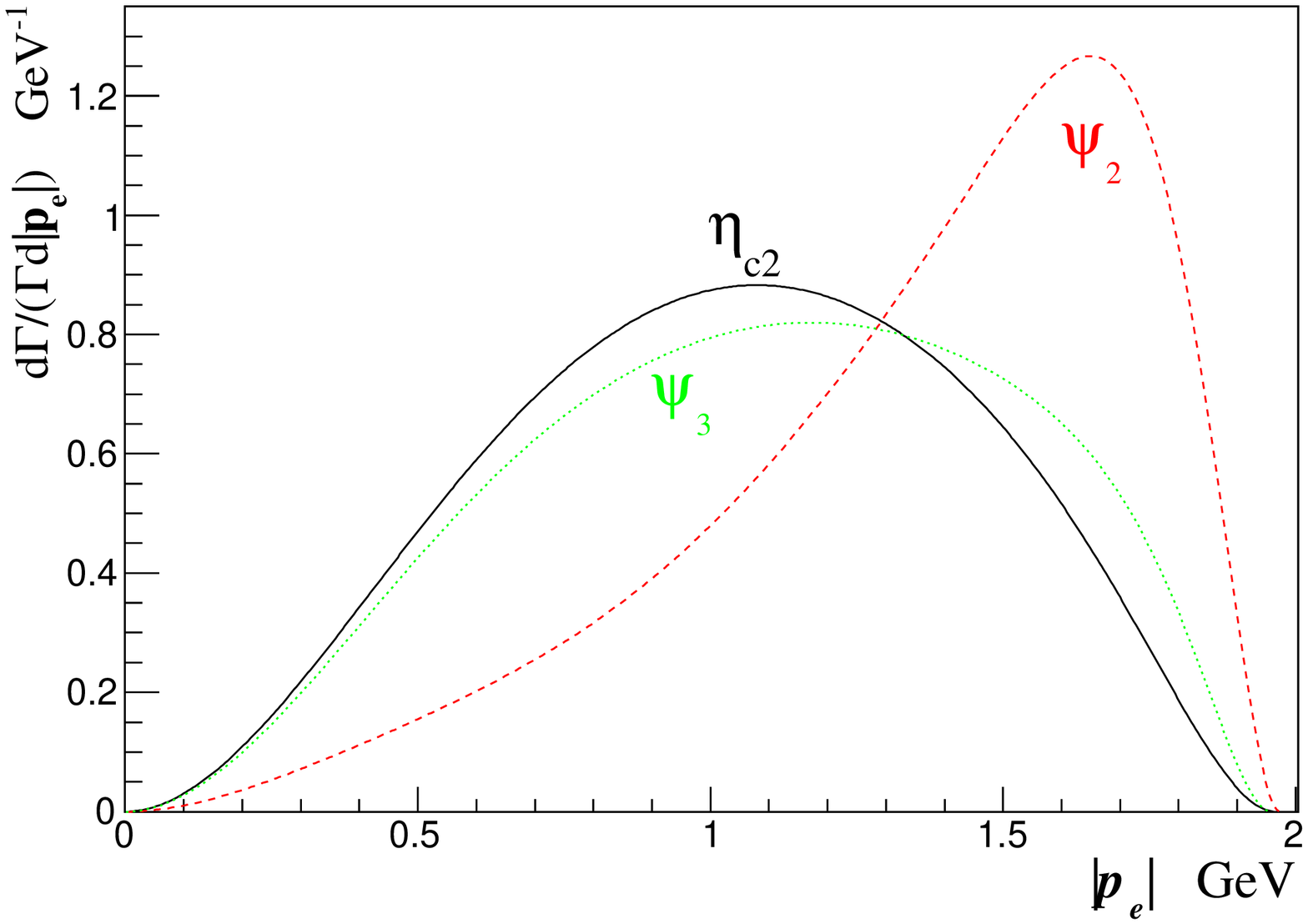} \label{Fig_dwe}}
\subfigure[Momentum spectrum for decay to $\tau$ mode.]   {\includegraphics[width=0.48\textwidth]{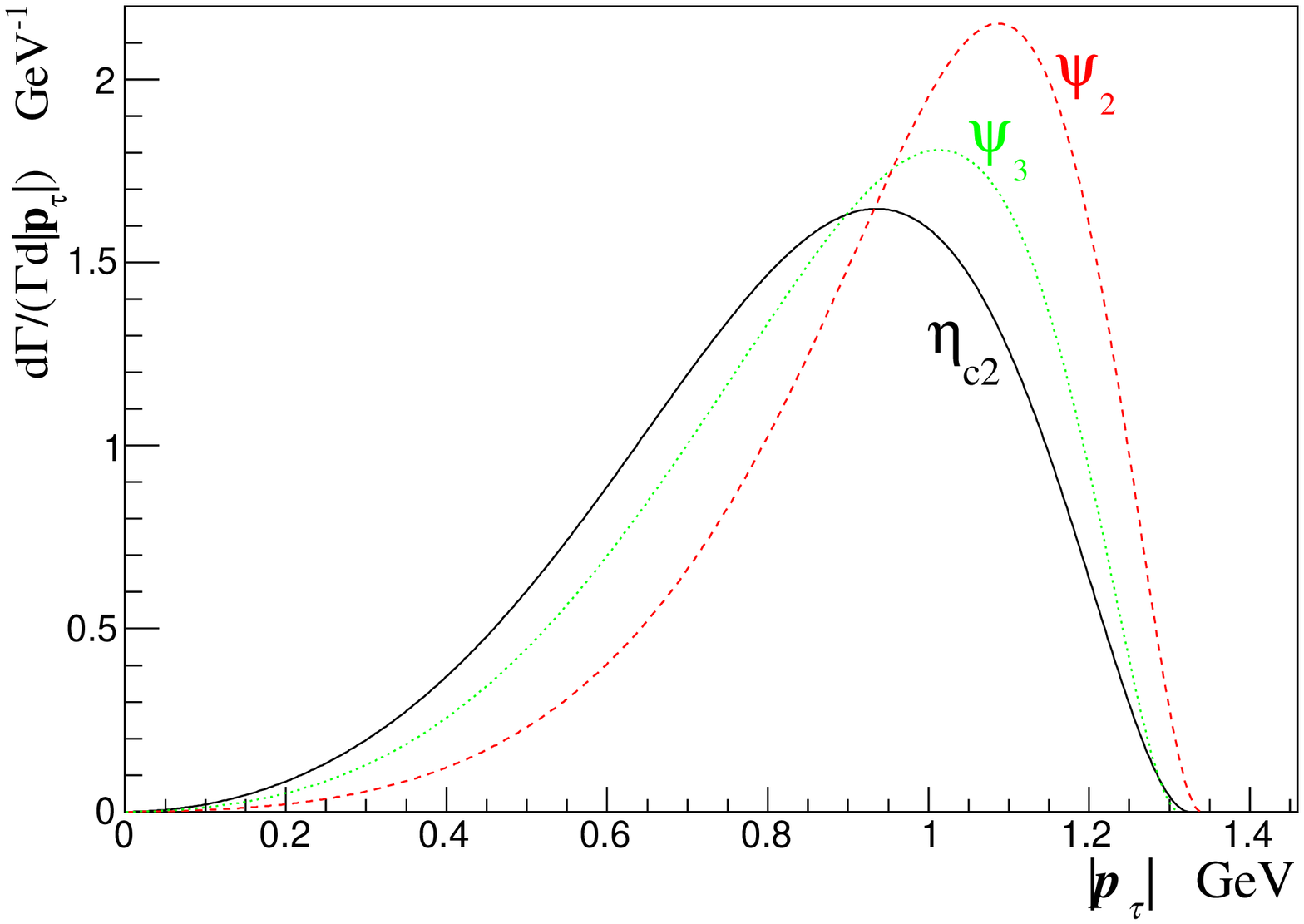} \label{Fig_dwt}}
\caption{The spectra of relative width vs charged leptons 3-momentum in $B_c$ semi-leptonic decays into $D$ wave charmonia. $|\bm{p}_e|$ and $|\bm{p}_\tau|$ are the 3-momentum amplitudes of $e$ and $\tau$, respectively. }\label{Fig-momentum}
\end{figure}

For the sake of completeness, we also plot~\autoref{Fig-cos} and~\autoref{Fig-momentum} to show the spectra of decay widths varying along $\cos \theta$ and 3-momentum ${|\bm p_\ell|}$ of the charged lepton, respectively. Here we do not give the result of $\mu$ mode which is almost the same as that of $\ell=e$. For the angular distribution in~\autoref{Fig-cos}, we can see when $\ell=e$, $\text{d}\Gamma/(\Gamma \text{d}\cos\theta)$ decreases monotonously for $\psi_2$ when $\cos\theta$ varies from $-1$ to 1, but reaches the maximum value for $\eta_{c2}$ and $\psi_3$ in the vicinity of 0. When $\ell=\tau$, all the three distributions are monotonic functions (for $\eta_{c2}$ and $\psi_3$, the angular spectra are increasing functions, while for $\psi_2$, it's a decreasing function). As to the momentum distribution (see~\autoref{Fig-momentum}), one can see the results of $\eta_{c2}$ and $\psi_3$ are more symmetrical than that of $\psi_2$, especially for $\ell=e$.  These results will be useful to the future experiments.

%\vspace{0.5em}
%\vspace{0.5em}

\subsection{Results of non-leptonic decays and uncertainties estimation}
The non-leptonic decay width of $B_c$ to $D$-wave charmonia are list in \autoref{nonW}. In the calculation, the decay constants of the charged mesons are~\cite{PDG-2014,PRD73-2006}
\begin{align*}
f_\pi=130.4 ~\si{MeV}, \quad f_K=156.2~\si{MeV}, \quad f_\rho=210~\si{MeV}, \quad f_{K^*}=217~\si{MeV}.
\end{align*}
The factorization method is used and the decay widths are expressed with general Wilson coefficient $a_1$. In this paper, to calculate the branching ratios of non-leptonic decays we choose $a_1=1.14$~\cite{PRD73-2006}.

\begin{table}[ht]
\caption{Non-leptonic decays width for $B_c^-$ to $\eta_{c2},~\psi_2$ and $\psi_3$ with general Wilson coefficient $a_1$.}\label{nonW}
\vspace{0.2em}
\centering
\begin{tabular}{lclclc}
&&&&&\hfill{$\times a_1^2$~(GeV) }\\
\toprule[1.5pt]
Channels~                  &Width~ ~                      &       Channels       		  &       Width~               &       Channels   	    & Width     \\
\midrule[1.2pt]
$B_c^-\To{\psi_2} \pi^- $   &         $1.2^{-0.1}_{+0.1}\e{-17} $         & $B_c^-\To{\eta_{c2}} \pi^- $   &       $4.4^{-0.6}_{+0.7}\e{-16} $          &$B_c^-\To{\psi_3} \pi^- $    & $1.9\er{0.3}{0.4}{-16} $   \vspace{0.3em} \\
$B_c^-\To{\psi_2} K^-   $   &         $8.3^{-0.7}_{+0.7}\e{-19} $        & $B_c^-\To{\eta_{c2}}  K^-  $   &      $3.2^{-0.4}_{+0.5}\e{-17} $         &$B_c^-\To{\psi_3} K^-   $    & $1.3\er{0.2}{0.3}{-17} $     \vspace{0.3em}\\
$B_c^-\To{\psi_2} \rho^-$   &         $1.1^{-0.1}_{+0.2}\e{-16} $          & $B_c^-\To{\eta_{c2}}\rho^- $   &     $9.1^{-1.0}_{+2.0}\e{-16} $          &$B_c^-\To{\psi_3} \rho^-$    & $4.6\er{0.7}{0.9}{-16} $   \vspace{0.3em}\\
$B_c^-\To{\psi_2} K^{\ast-} $   &     $7.1^{-0.9}_{+1.0}\e{-18} $         & $B_c^-\To{\eta_{c2}}K^{\ast-}  $   &  $4.8^{-0.7}_{+0.8}\e{-17} $         &$B_c^-\To{\psi_3} K^{\ast-} $    & $2.5\er{0.4}{0.5}{-17} $ \vspace{0.1em}   \\
\bottomrule[1.5pt]
\end{tabular}
\end{table}

The branching ratios of the non-leptonic decays are listed in \autoref{nonB} and \autoref{nonB-2}. For the channels with $\psi_2$ as the final charmonium, when the light meson is pseudoscalar, the branching ratio is smaller than that of Ref.~\cite{PRD73-2006} but about 20 times larger than that of Ref.~\cite{PRD74-2006}. While for the channels with vector charged mesons, the branching ratios are about 2 times and 5 times larger than those of Ref.~\cite{PRD73-2006} and Ref.~\cite{PRD74-2006}, respectively. Within all non-leptonic channels, those with $\rho$ as the charged meson have the largest branching ratios, which have more possibilities to be discovered by the future experiments.

\begin{table}[ht]
\caption{Branching ratios of non-leptonic decays for $B_c^-$ to $\psi_2$. $a_1=1.14$ and $\tau_{B_c}=0.452\times 10^{-12}$ \si{s}. The first uncertainties are from varying the model parameters by $\pm5\%$ then finding the maximum deviation. The second uncertainties are from the calculations of Wilson coefficient $a_1=c_1+\frac{1}{N_c}c_2$, where we change $N_c$ from 2 to $+\infty$ to estimate the non-factorizable contributions.}\label{nonB}
\vspace{0.2em}\centering
\begin{tabular}{lccc}
\toprule[1.5pt]
       Channels              &BR      			    &Ref.~\cite{PRD73-2006}             &Ref.~\cite{PRD74-2006}      \\
\midrule[1.2pt]
$B_c^-\To {\psi_2} \pi^- $   & $1.0\er{0.1-0.2}{0.1+0.4}{-5}$     &      $1.7\e{-5}$                  & $ 4.1\er{0.02}{0.03}{-7}$     \vspace{0.4em}\\
$B_c^-\To {\psi_2} K^-   $   & $7.4\er{0.6-1.4}{0.6+3.1}{-7}$     &      $1.2\e{-6}$                  & $ 3.1\er{0.2}{0.2}{-8}$      \vspace{0.4em}\\
$B_c^-\To {\psi_2} \rho^-$   & $9.6\er{1.0-1.7}{1.0+4.0}{-5}$     &      $5.5\e{-5}$                  & $ 2.0^{-0.3}\e{-5}$     \vspace{0.4em}\\
$B_c^-\To {\psi_2} K^{\ast-}$& $6.4\er{0.8-1.2}{1.0+2.7}{-6}$     &      $3.2\e{-6}$                  & $ 1.4^{-0.2}\e{-6}$     \vspace{0.1em}\\
\bottomrule[1.5pt]
\end{tabular}
\end{table}
\vspace{1ex}    %\vbox{}

\begin{table}[ht]
\caption{Branching ratios of non-leptonic decays for $B_c^-$ to $\eta_{c2}$ and $\psi_3$. $a_1=1.14$ and $\tau_{B_c}=0.452\times 10^{-12}$ \si{s}. See caption of~\autoref{nonB} for further explanations. }\label{nonB-2}
\vspace{0.2em}\centering
\begin{tabular}{lclc}
\toprule[1.5pt]
   Channels                     & BR      &       Channels             & BR     \\
\midrule[1.2pt]
$B_c^-\To {\eta_{c2}} \pi^- $   & $3.9\er{0.5-0.7}{0.6+1.6}{-4}$       &$B_c^-\To {\psi_3} \pi^- $     & $ 1.7\er{0.3-0.3}{0.3+0.7}{-4}$ \vspace{0.4em}\\
$B_c^-\To {\eta_{c2}}  K^-  $   & $2.8\er{0.4-0.5}{0.5+1.2}{-5}$    	  &$B_c^-\To {\psi_3} K^-   $     & $ 1.2\er{0.2-0.2}{0.2+0.5}{-5}$ \vspace{0.4em}\\
$B_c^-\To {\eta_{c2}}\rho^- $   & $8.1\er{1.0-1.5}{1.0+3.4}{-4}$   	  &$B_c^-\To {\psi_3} \rho^-$     & $ 4.1\er{0.7-0.7}{0.8+1.7}{-4}$ \vspace{0.4em}\\
$B_c^-\To {\eta_{c2}}K^{\ast-}$ & $4.3\er{0.6-0.8}{0.7+1.8}{-5}$       &$B_c^-\To {\psi_3} K^{\ast-}$  & $ 2.3\er{0.4-0.4}{0.5+0.9}{-5}$ \vspace{0.1em}\\
\bottomrule[1.5pt]
\end{tabular}
\end{table}
\vspace{1ex}    %\vbox{}

In order to estimate the systematic theoretical uncertainties for non-leptonic decays, we vary the parameters of Cornell potential model by $\pm5\%$ and then scanning the parameter-space to find the maximum deviation. From our results (see~\autoref{nonW}), the deviations of non-leptonic $B_c$ decays amount to $5\%\sim20\%$.

In the method of factorization approximation, the number of colors $N_c$, which appeared in the calculation of Wilson coefficient $a_1=c_1+\frac{1}{N_c}c_2$, is a parameter to be determined by experimental data. To estimate the systematic uncertainties from the non-factorizable contributions, we change the value of $N_c$ within the range $[2,+\infty]$, and then calculate the maximum deviation to the central values where $N_c=3$ and $a_1=1.14$ are used. In our calculations, this uncertainties can amount to about $15\%\sim 40\%$ in the non-leptonic decays of $B_c$ to $D$-wave charmonia, which are listed as the second uncertainties in the results of branching ratios in~\autoref{nonB} and~\autoref{nonB-2}.

\section{Summary}\label{Sec-5}
In this work we calculated semi-leptonic and non-leptonic decays of $B_c$ into the $D$-wave charmonia, namely, $\eta_{c2}(1^1\!D_2)$, $\psi_2(1^3\!D_2)$, and $\psi_3(1^3\!D_3)$, whose decay widths are expected to be narrow. The results show that for the semi-leptonic channels with the charged lepton to be $e$ or $\mu$, the branching ratios are of order of $10^{-4}$. For the non-leptonic decay channels, the largest branching ratio is also of order of $10^{-4}$. These results can be useful for the future experiments to study the $D$-wave charmonia.

\section*{Acknowledgments}
This work was supported in part by the National Natural Science
Foundation of China (NSFC) under Grant Nos.~11405037, 11575048 and 11505039, and in part by PIRS of HIT Nos.~T201405, A201409, and B201506. We thank Wei Feng of Bordeaux INP for her thorough proofreading of the manuscript.

\begin{appendix}

\section{Expressions for $N_i$s in the Hadronic Tensor $H_{\mu\nu}$}\label{Ni}

The hadronic tensor $N_i$ for $B_c$ to $^1\!D_2$ $c\bar c$ states are
\begin{align}
N_1 &= \frac{2 M^4 {p}_F^4 s_1^2}{3 M_F^4}-\frac{4 M^2 {p}_F^2 s_1 s_3}{3 M_F^2}-\frac{1}{2} M^2 {p}_F^2 s_4^2+\frac{s_3^2}{6}, \label{N1-2} \\
N_2 &= \frac{2 E_F M^3 {p}_F^2 s_1 s_3}{3 M_F^4}+\frac{E_F M^3 {p}_F^2 s_4^2}{2 M_F^2}-\frac{E_F M s_3^2}{6 M_F^2}+\frac{2 M^4 {p}_F^4 s_1 s_2}{3 M_F^4}-\frac{2 M^2 {p}_F^2 s_2 s_3}{3 M_F^2},\label{N2-2} \\
N_4 &= \frac{4 E_F M^3 {p}_F^2 s_2 s_3}{3 M_F^4}+\frac{2 M^4 {p}_F^4 s_2^2}{3 M_F^4}-\frac{M^4 {p}_F^2 s_4^2}{2 M_F^2}+\frac{M^2 s_3^2 (M_F^2+4 {p}_F^2)}{6 M_F^4}, \label{N4-2} \\
N_5 &= -\frac{M^4 {p}_F^4 s_4^2}{2 M_F^2}-\frac{M^2 {p}_F^2 s_3^2}{2 M_F^2}, \label{N5-2} \\
N_6 &= -\frac{M^2 {p}_F^2 s_3 s_4}{M_F^2}. \label{N6-2}
\end{align}
For $B_c$ to $^3\!D_2$ state the relations between $N_i~(i=1,2,4,5,6)$ and form factors $t_j~(j=1,2,3,4)$ are the same with $^1\!D_2$ state, just $s_j$ are replaced with $t_j$.

The hadronic tensor $N_i~(i=1,2,4,5,6)$ for $B_c$ to $^3\!D_3$ charmonium are expressed with corresponding form factors $h_j~(j=1,2,3,4)$ as
\begin{align}
N_1 &=  \frac{2 M^6 {p}_F^6 h_1^2}{5 M_F^6}-\frac{4 M^4 {p}_F^4 h_1 h_3}{5 M_F^4}-\frac{4 M^4 {p}_F^4 h_4^2}{15 M_F^2}+\frac{2 M^2 {p}_F^2 h_3^2}{15 M_F^2}, \label{N1-3}\\
N_2 &=  \frac{2 {E_F} M^5 {p}_F^4 h_1 h_3}{5 M_F^6}+\frac{4 {E_F} M^5 {p}_F^4 h_4^2}{15 M_F^4}-\frac{2 {E_F} M^3 {p}_F^2 h_3^2}{15 M_F^4}+\frac{2 M^6 {p}_F^6 h_1 h_2}{5 M_F^6}-\frac{2 M^4 {p}_F^4 h_2 h_3}{5 M_F^4}, \label{N2-3}\\
N_4 &=  \frac{4 {E_F} M^5 {p}_F^4 h_2 h_3}{5 M_F^6}+\frac{2 M^6 {p}_F^6 h_2^2}{5 M_F^6}-\frac{4 M^6 {p}_F^4 h_4^2}{15 M_F^4}+\frac{2 M^4 {p}_F^2 h_3^2 (M_F^2+3 {p}_F^2) }{15 M_F^6}, \label{N4-3}\\
N_5 &=  -\frac{4 M^6 {p}_F^6 h_4^2}{15 M_F^4}-\frac{4 M^4 {p}_F^4 h_3^2}{15 M_F^4}, \label{N5-3}\\
N_6 &=-\frac{8 M^4  {p}_F^4 h_3 h_4}{15 M_{\!F}^4}. \label{N6-3}
\end{align}.

\section{Expressions for $x_i$ in Form Factors $s_i$}\label{xi}
The expressions for $x_i~(i=1,2,\cdots,11)$ in  Eq.~(\ref{si-1}) are as below
\begin{align}
{x_1}&=-\frac{4 \alpha'^2_2 E_F^2}{M^4 M_F^2} {(\alpha'_2 {A_1}{B_4} E_F^2 M+{A_1} {B_1} M M_F^2+{A_3} {B_2} M_F \bm P_F \!\cdot \! \bm q + \alpha'_2 {A_4} {B_4} E_F \bm P_F \!\cdot \! \bm q )}. \label{x_1}\\
{x_2}&=+\frac{4 \alpha'^2_2 E_F^2}{M^3 M_F^2} {( \alpha'_2 {A_1} {B_4} E_F M-{A_2} {B_2} M M_F-{A_4} {B_4} {\bm q^2})}. \label{x_2}\\
{x_3}&=+\frac{4 \alpha'^2_2 E_F^2}{M^3 M_F^2} {({A_1} {B_4} E_F M-{A_3} {B_2} E_F M_F+{A_4} {B_1} M_F^2+{A_4} {B_4} \bm P_F \!\cdot \! \bm q )}. \label{x_3}\\
{x_4}&=+\frac{8 \alpha'_2 E_F}{M^3 M_F^2} {(\alpha'_2{A_1} {B_4} E_F^2 M+{A_1} {B_1} M M_F^2+{A_3} {B_2} M_F \bm P_F \!\cdot \! \bm q +\alpha'_2 {A_4}{B_4} E_F \bm P_F \!\cdot \! \bm q )}. \label{x_4}\\
{x_5}&=-\frac{8 \alpha'_2 E_F}{M^2 M_F^2} {( \alpha'_2{A_1} {B_4} E_F M-{A_2} {B_2} M M_F-{A_4} {B_4} {\bm q^2})}. \label{x_5}\\
{x_6}&=-\frac{8 \alpha'_2 E_F}{M^2 M_F^2} {({A_1} {B_4} E_F M-{A_3} {B_2} E_F M_F+{A_4} {B_1} M_F^2+{A_4} {B_4} \bm P_F \!\cdot \! \bm q) }. \label{x_6}\\
{x_7}&=-\frac{4}{M^2 M_F^2} {( \alpha'_2{A_1} {B_4} E_F^2 M+{A_1} {B_1} M M_F^2+{A_3} {B_2} M_F \bm P_F \!\cdot \! \bm q + \alpha'_2 {A_4} {B_4} E_F \bm P_F \!\cdot \! \bm q )}. \label{x_7}\\
{x_8}&=+\frac{4}{M M_F^2} {(\alpha'_2 {A_1} {B_4} E_F M-{A_2} {B_2} M M_F-{A_4} {B_4} {\bm q^2})}. \label{x_8}\\
{x_9}&=+\frac{4}{M M_F^2} {({A_1} {B_4} E_F M-{A_3} {B_2} E_F M_F+{A_4} {B_1} M_F^2+{A_4} {B_4} \bm P_F \!\cdot \! \bm q )}. \label{x_9}\\
{x_{10}}&=-\frac{8\alpha'_2 E_F }{M^3 M_F^2} {(-{A_1} {B_4} M+{A_3} {B_2} M_F+\alpha'_2 {A_4} {B_4} E_F)}. \label{x_10}\\
{x_{11}}&=+\frac{4}{M^2 M_F^2} {(-{A_1} {B_4} M+{A_3} {B_2} M_F+\alpha'_2 {A_4} {B_4} E_F)}. \label{x_11}
\end{align}
where $\alpha'_2=\frac{1}{2}$.

%---------------App-3
\section{BS positive wave function for $^3\!D_2$ and $^3\!D_3$ states}\label{fun-3d2-3d3}
The wave function for ${^3\!D_2}(2^{--})$ $c\bar c$ can be written as~\cite{3D23-wave}
\begin{equation} \label{wave-f2}
\varphi^{++}(^3\!D_2)=\mathrm{i}\epsilon_{\mu \nu \alpha \beta }\frac{P_F^{\nu}}{M_F}q'^{\alpha}_{\perp} e^{\beta \delta} q'_{\perp \delta} \gamma^{\mu} \bigg[ i_1+i_2 \frac{\dsl{P}{F}}{M_F} +i_4\frac{{\slashed P}\!_F {\slashed q'_\perp}}{M_F^2}  \bigg ].
\end{equation}
$i_1,~i_2$ and $i_4$ are defined as
\begin{equation}\label{par-f2}
\begin{aligned}
i_1=& \frac{1}{2} \biggl[I_1-\frac{\omega_c}{m_c}I_2  \bigg]    ,\\
i_2=& \frac{1}{2} \biggl[I_2-\frac{m_c}{\omega_c}I_1  \bigg]    ,\\
i_4=&-\frac{M_F}{\omega_c}i_1.
\end{aligned}
\end{equation}
$I_1$ and $I_2$ are functions of $q'^2_\perp$.

The positive part of the wave function of ${^3\!D_3}(3^{--})$ state has the form~\cite{3D23-wave}
\begin{equation} \label{wave-f3}
\varphi^{++}(^3\!D_3)=e_{\mu \nu \alpha} q^{\prime\nu}_{\perp} q_{\perp}^{\prime\alpha} \biggl [ q^{\prime\mu}_{\perp}( u_1+u_3 \frac{\slashed q'_{\perp}}{M_F} +u_4\frac{\dsl{P}{F} \slashed q_{\perp}}{M_F^2})
+ \gamma^\mu ( u_5 M_F + u_6 \dsl{P}{F}) +  u_8  \frac{(\gamma^\mu \dsl{P}{F} \slashed q'_{\perp} +  \dsl{P}{F} q'^\mu_{\perp})}{M_F}\bigg],
\end{equation}
where $u_i~(i=1,3,4,5,6,8)$ are expressed as
\begin{equation}\label{par-f3}
\begin{aligned}
u_1=& \frac{\omega_c(q_{\perp}^2U_3+M_F^2U_5)+m_c(q_\perp^2U_4-M_F^2 U_6)}{2M_F m_c \omega_c},\\
u_3=&\frac{1}{2}\biggl[U_3+\frac{m_c}{\omega_c}U_4-\frac{M_F^2}{m_c\omega_c}U_6 \biggl],\\
u_4=&\frac{1}{2}\biggl[U_4+\frac{\omega_c}{m_c}U_3-\frac{M_F^2}{m_c\omega_c}U_5 \biggl],\\
u_5=& \frac{1}{2}\biggl[U_5-\frac{\omega_c}{m_c}U_6 \biggl],\\
u_6=& \frac{1}{2}\biggl[U_6-\frac{m_c}{\omega_c}U_5\biggl],\\
u_8=& -\frac{M_F}{\omega_c}u_5.
\end{aligned}
\end{equation}
In above expressions $U_3~,U_4,~U_5$ and $U_6$ are functions of $q'^2_\perp$, which could be determined numerically by solving the full Salpeter equation.

%--------------
\end{appendix}

\end{document}